\documentclass{jpp}
\usepackage{graphicx}

\usepackage[utf8]{inputenc}
\usepackage[T1]{fontenc}
\usepackage{amsmath}
\usepackage{xcolor}

\newcommand{\eq}[1]{equation~(\ref{#1})}
\newcommand{\eqs}[2]{equations~(\ref{#1}--\ref{#2})} 

\shorttitle{PIC simulations of tearing}
\shortauthor{K. M. Schoeffler et al.}

\title{Particle-in-cell simulations of the tearing instability for relativistic pair plasmas}

\author{K. M. Schoeffler\aff{1}
  \corresp{\email{Kevin.Schoeffler@rub.de}},
  B. Eichmann\aff{1},
  F. Pucci\aff{2},
  M. E. Innocenti\aff{1}}

\affiliation{\aff{1}Institut f\"ur Theoretische Physik, Ruhr-Universit\"at Bochum, Bochum, Germany\\
\aff{2}Jet Propulsion Laboratory, Pasadena, CA 91109, USA}

\begin{document}

\maketitle

\begin{abstract}
Two-dimensional particle-in-cell (PIC) simulations explore the collisionless tearing instability developing in a Harris equilibrium configuration in a pair (electron-positron) plasma, with no guide field, for a range of parameters from non-relativistic to relativistic temperatures and drift velocities. Growth rates match predictions of \cite{Zelenyi1979} modified for relativistic drifts by \cite{Hoshino2020} as long as the assumption holds that the thickness $a$ of the current sheet is larger than the Larmor radius $\rho_L$, with the fastest growing mode at $ka \approx 1/\sqrt{3}$. Aside from confirming these predictions, we explore the transitions from thick to thin current sheets and from classical to relativistic temperatures. We show that for thinner current sheets ($a< \rho_L$), the growth rate matches the prediction for the case $a=\rho_L$. We also explore the nonlinear evolution of the modes. While the wave number with the fastest growth rate initially matches the prediction of \cite{Zelenyi1979}, these modes saturate moving the dominant mode to lower wave numbers (especially for thick current sheets with low growth rates). Furthermore, at a late, non-linear stage, the growth rate (initially following the growth rate prediction proportional to $(\rho_L/a)^{3/2} < 1$) increases faster than exponentially, reaching a maximum growth rate equivalent to the linear growth rate prediction at $\rho_L/a = 1$, before eventually saturating.
\end{abstract}

\section{Introduction}\label{sec:intro}
When opposite-directed magnetic fields are separated by a thin current sheet (where either collisional or kinetic effects are present), the free energy of the magnetic field can be converted to perpendicular fields and bulk flows that further drive this process known as the tearing instability \citep{Furth1963}. The tearing instability corresponds to the initial stage of a process that can eventually develop into non-linear magnetic reconnection, and convert this free energy into more bulk flows, plasma heating, and non-thermal high-energy particles. On the other hand, competing instabilities i.e. kink \citep{Zenitani2007,Cerutti2014}, firehose \citep{Liu2012,Innocenti2015}, flow shears \citep{Faganello2010,Cassak2011}, or other nonlinear effects can, in some cases, disrupt or prevent the nonlinear stage of tearing from continuing.

The tearing instability has been studied for the last few decades, in several different regimes ranging from collisional tearing, which can be measured in the lab, and collisionless or very weakly collisional tearing, which is often the relevant regime in astrophysical and space plasmas \citep{Laval1966,Coppi1966}. Although not the focus of this paper, the plasma beta (ratio of magnetic pressure to plasma pressure) and the ratio of the guide field to the reconnecting field, can also play important roles in describing the tearing instability.

In extreme astrophysical environments \citep{Zenitani2007,Cerutti2014}, the magnetization $\sigma_c$, i.e. the ratio of the background magnetic field to the particle rest energy density, is much larger than $1$, and pair production often leads to a plasma predominantly composed of electrons and positrons.
Several works have studied the tearing instability in this context with analytical or numerical calculations of the growth rate in both kinetic and fluid regimes \citep{Zelenyi1979,Petri2007,Yang2017,Yang2019b,Yang2019}. Numerical studies have addressed the tearing instability using fluid models \citep{Komissarov2006,Barkov2016} and particle-in-cell (PIC) methods \citep{Zenitani2007,Bessho2007,Yin2008,Bessho2012,Liu2015,Zenitani2017}. However, to our knowledge, an extensive study of the tearing instability using PIC methods has not been offered. 

Here, we will present such a study, considering the high $\sigma_c$ pair plasma regime. We will therefore neglect the effects of the background plasma, and consider a mass ratio of $1$. Pair plasmas are produced in environments with extremely high energy density, so temperatures are expected to be relativistic. On the other hand, pairs can be strongly cooled by radiative processes, allowing for classical temperatures as well. We will therefore consider a wide range of temperatures. Out of simplicity, we will consider a fully collisionless Harris equilibrium~\citep{Harris1962} configuration with no guide field.
Also, we note that asymmetric reconnection has been studied in similar contexts \citep{Mbarek2022}, but we will consider a symmetric configuration.

With these assumptions, the problem reduces simply to 2 parameters, the temperature of the plasma normalized to the electron rest mass energy $T/m_e c^2$, which is the same for electrons and positrons, and the ratio of the Larmor radius (based on the upstream magnetic field) to the thickness of the current sheet $\rho_L/a$.
A third parameter that is a function of the other two is the proper drift velocity compared to the speed of light, $u_d/c$. 

A quite general theoretical model for this instability that is relevant for all the assumptions that we are considering was derived in \cite{Zelenyi1979}. The study included limits valid for both non-relativistic
$T/m_e c^2 \ll 1$ and ultra-relativistic $T/m_e c^2 \gg 1$ regimes, but assumed that $\rho_L/a \ll 1$, i.e. a thick current sheet with respect to the kinetic scales involved in supporting the reconnection process. (Note that these current sheets are often considered thin with respect to the system size as addressed in Section~\ref{sec:astrolim}.) This assumption implies $u_d/c \ll 1$. However, a recent paper \citep{Hoshino2020} extends the model beyond this constraint in the relativistic temperature regime. He shows with both theory and empirically through simulation results that Zelenyi's model with an additional factor of $1/\Gamma_d$,
where $\Gamma_d = \sqrt{1+u_d^2}$ is a good prediction of the growth rate even for $u_d/c \gg 1$, resulting in a maximum growth rate for the tearing instability at $u_d/c \sim 1$. In this paper, we show using PIC simulations that Zelenyi's model, including Hoshino's extension, gives quite accurate results for a wide range of parameters. While there are modifications to the theoretical model based on the mass ratio for electron-ion plasmas included in Zelenyi's model, which are beyond the scope of this paper, the electron-positron solution gives a good order of magnitude estimation of the growth rate even in those situations.

We now lay out the organization of the paper. After this introduction in Section~\ref{sec:intro}, we will describe our setup of the simulation, the Harris equilibrium, and important length scales of the problem in Section~\ref{sec:setup}. We will then describe the equations from Zelenyi's model in Section~\ref{sec:theory}. We explain our simulation results in Section~\ref{sec:results} which is divided into two subsections; one for a set of runs with classical parameters and one for a set with relativistic parameters. In Section~\ref{sec:astrolim} we explore limits on astrophysical configurations based on the theoretical model. Finally, we will conclude with a discussion in Section~\ref{sec:results}.
 
\section{Simulation Setup}\label{sec:setup}
The simulations presented here begin in a double Harris equilibrium using the relativistic generalization ~\citep{KirkHarris} for relativistic temperatures ($T > m_e c^2/2$) with periodic boundary conditions. We use a simulation box ranging from $x = -L_x$ to $L_x$, and $y = -L_y$ to $L_y$, where $L_y$ is the distance between the two current sheets.  

The current and self-consistent magnetic field profiles are in pressure balance in a kinetic equilibrium.
The current is carried by counter-drifting Maxwellian or Maxwell-J\"uttner distributions of positrons and electrons with a uniform temperature~$T$, boosted into opposite $\pm \hat{z}$-directions with a uniform velocity~$v_d$.  
The lab-frame density profile (of both electrons and positrons) in the Harris current sheet at $y = \pm L_y/2$ is:
\begin{equation}
	n = \frac{n_0}{2} {\rm sech}^2 \left(\frac{y \mp L_y/2}{a}\right),
\end{equation}
where $n_0$ is the total (electron plus positron) density at the center of each current sheet. 

The self-consistent initial reconnecting magnetic field is:
\begin{equation}
	B_x =B_0\left[
	1 -\tanh\left(\frac{y - L_y/2}{a}\right)
	+\tanh\left(\frac{y + L_y/2}{a}\right)\right].
\end{equation}

Note that we do not consider a background population $n_b$; this assumption corresponds to the limit where $\sigma_c = B_0^2/4 \pi n_b m_e c^2 \gg 1$.

The drift velocity $v_d$ corresponds to a Lorentz factor
$\Gamma_d = 1/\sqrt{1-v_d^2/c^2}$, and a proper drift velocity $u_d = \Gamma_d v_d$. The magnetic field can be calculated, using pressure equilibrium, to be
\begin{equation}
\label{forcebalance}
    B_0 = \sqrt{\frac{8 \pi n_0 T}{\Gamma_d}},
\end{equation}
and using Ampere's law, the current half-thickness can be calculated to be
\begin{equation}
\label{ampere}
    a = \frac{c B_0}{4 \pi e n_0 v_d} = \sqrt{\frac{T c^2}{2 \pi n_0 e^2 \Gamma_d v_d^2 }} \approx \sqrt{\frac{\Gamma_T m_e c^4}{4 \pi n_0 e^2 \Gamma_d v_d^2 }}.
\end{equation}
As highlighted in \cite{Pucci2018}, tearing growth rates can be affected by the communication between two nearby (i.e. when $L_y/a$ is small) current sheets. Based on the analysis of our simulations, $L_y/a\approx 20$ is a sufficient distance to guarantee no interaction between the current sheets. We therefore adopt this separation in all of the simulations presented.
The constraints from \eqs{forcebalance}{ampere} leave only two free parameters, $T$ and $u_d$ (as we do not consider collisions or radiation, $n_0$ can be absorbed into the normalization).  In the relativistic regime, we will write expressions in terms of the peak Lorentz factor in a static, but strongly relativistic Maxwell-J\"uttner distribution $\Gamma_T \equiv 2T/m_e c^2$, which is simply a function of the temperature. Likewise, in the classical regime, we will write expressions in terms of the thermal velocity $v_T$ ($v_T/c \equiv \sqrt{2T/m_e c^2} $).

We can express the scales of the system in terms of these free parameters:
the classical electron inertial length:
\begin{equation}
\label{classde}
    d_{e,C} = \sqrt{\frac{m_e c^2}{4\pi n_0 e^2}},
\end{equation}
the relativistic electron inertial length:
\begin{equation}
\label{relde}
    d_{e,R} = \sqrt{\Gamma_T} d_{e,C} ,
\end{equation}
the classical Larmor radius:
\begin{equation}
\label{classrho}
    \rho_{L,C} = \frac{v_T}{\Omega_c} = \sqrt{\Gamma_d} d_{e,C} = \frac{u_d}{v_T}a,
\end{equation}
and the relativistic Larmor radius:
\begin{equation}
\label{relrho}
    \rho_{L,R} = \frac{\Gamma_T c}{\Omega_c} = \sqrt{\Gamma_d} d_{e,R} = \frac{u_d}{c}a,
\end{equation}
where $\Omega_c = e B_0/m_e c$ is the cyclotron frequency. Our constraint from force balance, \eq{forcebalance}, implies $\rho_{L} \approx d_e$ in both classical and relativistic regimes
as seen in \eqs{classrho}{relrho} as long as $\Gamma_d\sim1$. We do not precisely define $\rho_L$ in the transition between the classical and relativistic regimes, at $T/m_ec^2 \sim 1$ when $\rho_{L,C} \sim \rho_{L,R}$. We will therefore specify in the text when we are using $\rho_{L,C}$ or $\rho_{L,R}$.

\section{Theoretical model}\label{sec:theory}
\cite{Zelenyi1979} calculates a growth rate for the tearing instability in a non-relativistic and ultra-relativistic Harris sheet assuming a small $\rho_L/a$. The classical growth rate assuming a pair plasma with equal mass $m_e$ and equal temperature $T$ is
\begin{equation}
\label{classicaltear}
        \frac{\gamma a}{c} \approx \frac{1}{\sqrt{\pi}} k a \left(1-k^2a^2\right) \left(\frac{u_d}{c}\right)^{3/2} \left(\frac{v_{T}}{c}\right)^{-1/2}
\end{equation}
and the fully relativistic case
\begin{equation}
\label{reltear}
\frac{\gamma a}{c} \approx \frac{2\sqrt{2}}{\pi} k a \left(1-k^2a^2\right) \frac{1}{\Gamma_d^{5/2}} \left(\frac{u_d}{c}\right)^{3/2},
\end{equation}
where we have added the factor of $1/\Gamma_d^{5/2}$ (or $1/\Gamma_d$, if you write the equation in terms of the drift velocity $v_d/c$) determined in \cite{Hoshino2020}.
Note that both growth rates have a maximum growth rate at the wave number $ka = 1/\sqrt{3}$.

This prediction is based on the constant-$\psi$ approximation \citep{Burkhart1989}, which is valid in the limit that $ka \approx 1$. It is applicable down to values of $ka\sim k_{max}a$ corresponding to the maximum growth rate, below which the instability develops in the large $\Delta'$ regime (see e.g.~\citealt{DelSarto2016} and references therein). Therefore, the prediction that $ka \approx 1/\sqrt{3}$ is only an estimate. Several analytical as well as numerical methods, including PIC studies, have attempted to predict a more accurate dispersion relation \citep{Chen1985,Daughton1999,Daughton2003,Daughton2005,Petri2007}. In all studies, the wave number remains close to $k_{max}a \sim 1/2$, suggesting the results found in this paper are in agreement with the literature. In addition, this is consistent with the idea that the simulation predicted wave-vector is at the transition between the constant-$\psi$ and regime of the maximum growth rate (see Fig.~4 \citealt{2016JPlPh..82e5301T} for the resistive tearing case, and Fig.~1 \citealt{DelSarto2016} for the collisionless case).

Using \eq{classrho} we can write \eq{classicaltear} in terms of $\rho_{L,C}/a$:
\begin{equation}
\label{classicaltear2}
        \frac{\gamma a}{c} \approx \frac{1}{\sqrt{\pi}} k a \left(1-k^2a^2\right) \left(\frac{u_d}{v_T}\right)^{3/2} \frac{v_T}{c}= \frac{1}{\sqrt{\pi}} k a \left(1-k^2a^2\right) \left(\frac{\rho_{L,C}}{a}\right)^{3/2} \frac{v_T}{c}.
\end{equation}
This is equivalent to predictions from \cite{Laval1966} and \cite{Coppi1966}, except for the numerical factors and $k$ dependence.
We can similarly combine \eq{relrho} and \eq{reltear} to show that $\gamma \sim  (\rho_L/a)^{3/2}$ for both classical and relativistic regimes.

In the next section, we will test Zelenyi's model for the non-relativistic regime using the previous equation for constant values of $\rho_{L,C}/a$, which he assumes to be small, as a function of the temperature $T = m_e v_T^2/2$.
We will also explore the $T$, $u_d$ space from $T \ll m_e c^2$ to $T \gg m_e c^2$ again testing Zelenyi's model. We should note that the model is only valid for sufficiently large temperatures. For increasingly smaller temperatures (and constant $u_d$), $\rho_{L,C}/a = u_d/v_T$ will increase until $\rho_{L,C}/a \sim 1$, and the assumptions of the model break down.

While we consider the case with no constant guide field pointed in the direction perpendicular to the plane of the simulation, \cite{Zelenyi1979} also discussed a regime where the guide field magnetizes the particles at all points in space. In the present paper, we will not investigate this regime, but it is worth noting and comparing it with other models. In this regime, the growth rate is proportional to $(\rho_L/a)^2$ rather than $(\rho_L/a)^{3/2}$. This matches other kinetic studies like \cite{Drake1977} as well as fluid models such as \cite{DelSarto2016}, \cite{Betar2022}, etc. who find growth rates that depend on $(d_e/a)^2$ in the respective small $\Delta^\prime$ EMHD and RMHD (where $\rho_{L} \sim d_e$) regimes. As we showed in the previous section, force balance implies that $\rho_{L} \approx d_e$ for the Harris equilibrium with no guide field. Other models exist where $d_e \gg \rho_{L}$ are only valid in regimes either with strong guide fields or starting from an equilibrium that differs from a Harris sheet, e.g. a force-free condition.

\section{Simulation results}\label{sec:results}
In this section, we test the theories for the classical cases where the temperature remains nonrelativistic ($T/m_ec^2 \ll 1$), and in the more general case including relativistic temperatures using particle-in-cell simulations, taking advantage of the OSIRIS framework~\citep{OSIRIS}.
\subsection{Classical tearing}
Here we present results from simulations aimed at measuring the tearing growth rate and verifying \eq{classicaltear2}. We note that, unlike classical references \cite{Laval1966} and \cite{Coppi1966}, we are considering the case of a pair plasma composed of positrons and electrons with equal mass. We expect pair plasmas with non-relativistic temperatures to occur as a result of radiative cooling. 
Furthermore, our general conclusions should be relevant for electron-proton plasmas as well, as the predictions of the growth rate from \cite{Zelenyi1979} with electron-proton mass ratios only differ by a factor of order unity (as long as the temperature ratio also remains of order unity).
We will examine two regimes holding $a/\rho_{L,C} = 2.5$ and $a/\rho_{L,C} = 5$ constant, and varying the temperature, in the regime where $T/m_ec^2 \ll 1$. This means that we are also varying $u_d/c$, in contrast with the next section where we will hold $a/\rho_{L,R} = 1/(u_d/c)$ constant. Please note the different usage of classical and relativistic Larmor radii, $\rho_{L,C}$ and $\rho_{L,R}$ in the paragraph above.

For the cases with $a/\rho_{L,C} = 2.5$, we use 1024 particles-per-cell, $L_y/a = 20.5$ and $L_x = L_y/2$ with a resolution of 18.6 grid cells per $a$.
We take a time step of $dt = 0.035 a/c < dx/c/\sqrt{2} = 0.0376 a/c$ to satisfy the Courant condition.
For the case with $a/\rho_{L,C} = 5$, to avoid issues with numerical heating, we use 4096 particles-per-cell. We use the same system size, the same resolution of grid cells per $a$, and the same time step as for $a/\rho_{L,C} = 2.5$. The parameters of each of these runs can be found in Table \ref{table}.
\begin{table}
  \begin{center}
\def~{\hphantom{0}}
  \begin{tabular}{lcccccccc}
      $a/\rho_{L,C}$  & $T/m_e c^2$    &   $u_d/c$   & 
      $\gamma_{th} a/c$ & $\gamma_m a/c$ & $\gamma_m(L_x=L_y) a/c$& $t_{\rm{st,nl}}\gamma_{th}$ & $t_{\rm{fi,nl}}\gamma_{th}$& $\gamma_{m,nl} a/c$\\[3pt]
       2.5   & 0.0003125 & 0.01~  & 0.00137~& 0.00124~&-& 7.14& 7.36 & 0.0036\\
       2.5   & 0.00125~~ & 0.02~  & 0.00275~& 0.00277~&-& 7.69& 7.80& 0.0081\\
       2.5  & 0.005~~~~ & 0.04~  & 0.00549~& 0.00512~&-& 7.25& 7.47& 0.0139\\
       2.5   & 0.02~~~~~ & 0.08~ & 0.0110~~& 0.0108~~&- & 7.25& 7.47& 0.0282\\
       5   & 0.0003125 & 0.005  & 0.000486& 0.000465&0.000539& 5.31& 5.33& 0.0077\\
       5   & 0.00125~~ & 0.01~ & 0.000971& 0.000784&0.000961& 5.21& 5.22 & 0.0176\\
       5  & 0.005~~~~ & 0.02~& 0.00194~& 0.00150~&0.00194~ & 5.48& 5.52& 0.0297 \\
       5   & 0.02~~~~~ & 0.04~  & 0.00388~& 0.00366~&0.00277~& 4.86& 4.91& 0.0507\\
  \end{tabular}
  \caption{Parameters for the classical set of simulations, along with the theoretical linear growth rate $\gamma_{th}$ given by \eq{classicaltear2}, and the measured growth rate $\gamma_m$ using a best fit between $t \gamma_{th} = 3.08-4.39$ for cases with   $a/\rho_{L,C}=2.5$, and between $t \gamma_{th} = 1.55-3.88$ for cases with $a/\rho_{L,C}=5$ for standard simulations with $L_x = L_y/2$. For the simulations with $L_x = L_y$ the growth rate is measured after performing a low pass filter over the same time range. In addition, we include the time at the start $t_{\rm{st,nl}}$ and the finish $t_{\rm{fi,nl}}$ of the measurement of the fast-growing nonlinear growth rate $\gamma_{m,nl}$.
  }
  \label{table}
  \end{center}
\end{table}

\begin{figure}
\includegraphics[width=\textwidth]  {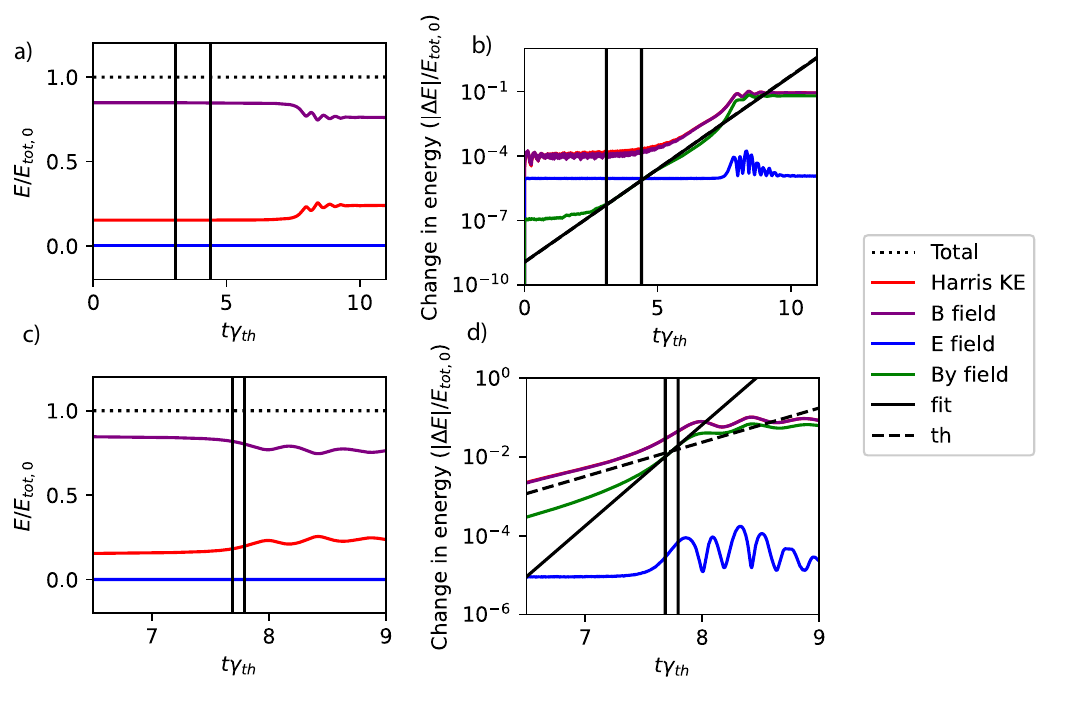}
  \caption{Evolution of the energy (a) and change of energy (b) in the Harris sheet electrons/positrons (kinetic energy), electric, and magnetic fields, as well as the $y$ component of the magnetic field that characterizes the tearing growth rate, for the simulation with $T/m_e c^2 = 0.0012$ and $a/\rho_{L,C} = 2.5$. A fit of growth is plotted in solid black along with the      theoretical growth rate, given by \eq{classicaltear2}, in the dashed line, which is nearly indistinguishable from the solid line.
  The same plots are also shown with a time range near the fast-growing nonlinear stage of the energy (c) and change of energy (d), where the fit for the faster growth rate is highlighted, and compared to the linear theoretical curve (for $a/\rho_{L,C} = 2.5$).
  The fits are measured in the range between the two vertical black lines.}
\label{fig:classicgrowth}
\end{figure}

We track the evolution of the perpendicular magnetic field energy $B_y^2/4\pi$ as a function of time, along with the kinetic energy of the particles in the Harris sheet, the total magnetic field energy, and the total electric field energy. In Figure \ref{fig:classicgrowth}, we present an example case where $T/m_ec^2 = 0.00125$ and $a/\rho_{L,C} = 2.5$. The magnetic energy $B_y^2/4\pi$ is dominated by noise up until $t \gamma_{th} \sim 2.5$ ($t c/a \approx 900$),  where we have normalized to the theoretical linear growth rate $\gamma_{th}$ given by \eq{classicaltear2}. This time, therefore, corresponds to a couple of e-folding times. We then measure a best fit of the growth rate between $t \gamma_{th} = 3.08-4.4$, obtaining a growth rate $\gamma_m a/c = 0.00277$, which matches very well with \eq{classicaltear2} ($\gamma_{th} a/c = 0.00275$). In the time interval between $t \gamma_{th} = 7.7-7.8$, the signal begins to grow faster ($\gamma a/c = 0.0081$), as measured in Figure \ref{fig:classicgrowth} (d). This faster growth rate is close to the linear prediction for $a/\rho_{L,C} = 1$ ($\gamma a/c = 0.011$). While the growth rate fits an exponential, it corresponds to multiple interacting modes, and we call this period the fast-growing nonlinear stage. Soon after the signal saturates, and a significant portion of the free energy of the magnetic field $B_x^2/4\pi$ is transferred to both the $B_y^2/4\pi$ signal and kinetic energy of the plasma, as seen in Figure \ref{fig:classicgrowth} (a,c), which shows the transfer of magnetic energy in purple to kinetic energy in red, and in Figure \ref{fig:classicgrowth} (b,d) which shows that $B_y$ in green constitutes a significant portion of the total magnetic energy in purple. The noise in the kinetic energy is larger than that of the $B_y^2/4\pi$ signal, so it is not useful for calculating a reliable slope. However, in Figure \ref{fig:classicgrowth} (b) at late times $t \gamma_{th} \sim 6.6$, the slope of the kinetic energy becomes comparable to that of $B_y^2/4\pi$.
The electric field energy $E_z^2/4\pi$ does not increase appreciably until the fully nonlinear stage.

We will now provide a potential explanation for the faster nonlinear growth stage that works in both relativistic and nonrelativistic regimes. 
In the nonlinear stage of the instability, the local $B_x$ decreases around the x-line effectively increasing $\rho_{L}$. This increase coincides with an increased ratio $\rho_{L}/a$  as long as $a$ does not grow too much, and simulations show that $a$, on the contrary, shrinks during this nonlinear stage.
We thus expect an increase in the instability growth rate from \eq{classicaltear2} or in relativistic cases \eq{reltear}, until $\rho_{L}/a \sim 1$, where the assumptions behind the derivation of \eqs{classicaltear}{classicaltear2} break down. For increasingly wide $\rho_{L}/a$ the growth rate will stop increasing with $\rho_{L}/a$ and begin to decrease; thus its maximal value should be at $\rho_{L}/a \sim 1$. We test this prediction in the classical and relativistic temperature regimes.
We will show in the relativistic part of Section \ref{sec:results} that when varying $T/m_ec^2$ (classical temperatures), keeping $\rho_{L,R}/a = u_d/c$ constant, the peak growth rate indeed occurs when $\rho_{L,C}/a \sim 1$.
We also provide evidence of a maximum when varying $u_d/c$ and keeping $T/m_ec^2$ constant.
While in the classical regime, a wider $\rho_{L,C}/a$ can occur if either $T/m_ec^2$ or $u_d/c$ change, in the relativistic regime, a wider $\rho_{L,R}/a$ implies a faster $u_d/c$. In particular, we expect the maximal value for relativistic cases where $\rho_{L,R}/a = u_d/c \sim 1$, because this is the maximal growth rate according to the predictions of \cite{Hoshino2020}. 
One should note that this is in a highly nonlinear stage, and thus linear growth rates can only be used as a rough estimate of the dynamics. On the other hand, we will show that this estimation gives a rather accurate prediction of the peak nonlinear growth rate.
As we showed in Section \ref{sec:setup}, force balance implies that $\rho_{L} \approx d_e \sim 1/\sqrt{n}$ (in both classical and relativistic regimes). In the regions where $B_x$ decreases, the density $n$ also decreases, and this force balance appears to hold. 
Following this logic, if there were a background population, the growth of $d_e$ would be limited to the background value $d_e(n_b)$, and the fastest nonlinear growth might also be limited to the prediction for $\rho_L/a = d_e(n_b)/a$ instead of $\rho_L/a = 1$. We check this prediction at the end of this section. 

\begin{figure}
  \centering
  \includegraphics[width=0.8\textwidth]  {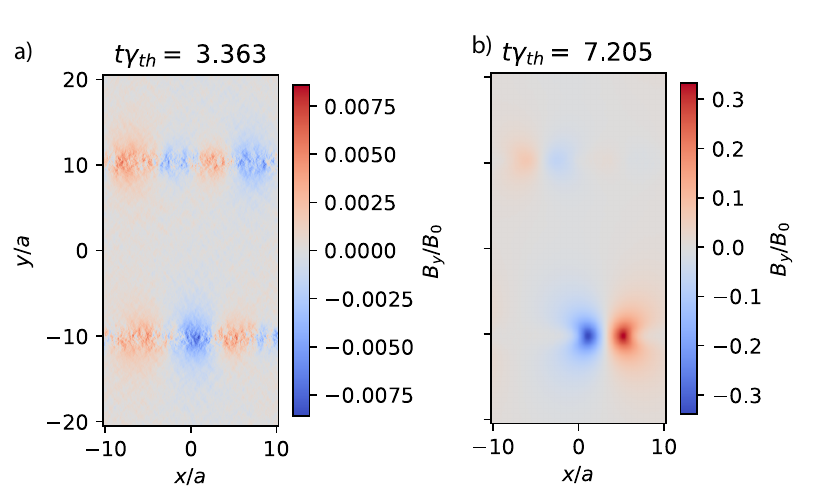}
  \caption{Map of $B_y$ as a function of space for the simulation with $T/m_e c^2 = 0.00125$ and $a/\rho_{L,C} = 2.5$ at an early time where the wave number $ka \approx 1/\sqrt{3}$ matches Zeleyni's prediction, and at a later time where the smallest $k$ ($m=1$) mode begins to dominate.}
\label{fig:classicby}
\end{figure}

In Figure \ref{fig:classicby} we show a map of the $B_y$ component of the magnetic field from the same example case from Figure \ref{fig:classicgrowth}, for two representative times. The first time at $t\gamma_{th}=3.363$ corresponds to the linear stage, where the signal has just grown beyond the noise. It is clear in the upper current sheet that the dominant mode is at $ka = 2\pi m a/(2 L_x) \approx 0.6$ ($m=2$). This matches very well with the predicted value from Zelenyi's model $ka = 1/\sqrt{3} \approx 0.58$.
The later time $t\omega_{pe}=7.205$ corresponds to a late stage of the linear growth rate, where the dominant mode shifts to a lower $k$ ($m=1$). Soon after, the growth moves into the fast-growing nonlinear stage. 
The start of the nonlinear stage matches with the prediction from \cite{Hoshino2021} based on the theory from \cite{Galeev1978}, that once $B_y/B_0 > k\rho_L$ an explosive nonlinear stage occurs. In our case, assuming $ka=1/\sqrt{3}$,  $k\rho_{L,C} \equiv 0.23$, which is on the same order as the $B_y/B_0$ seen in \ref{fig:classicgrowth}(b).

\begin{figure}
  \centering
  \includegraphics[width=0.7\textwidth]  {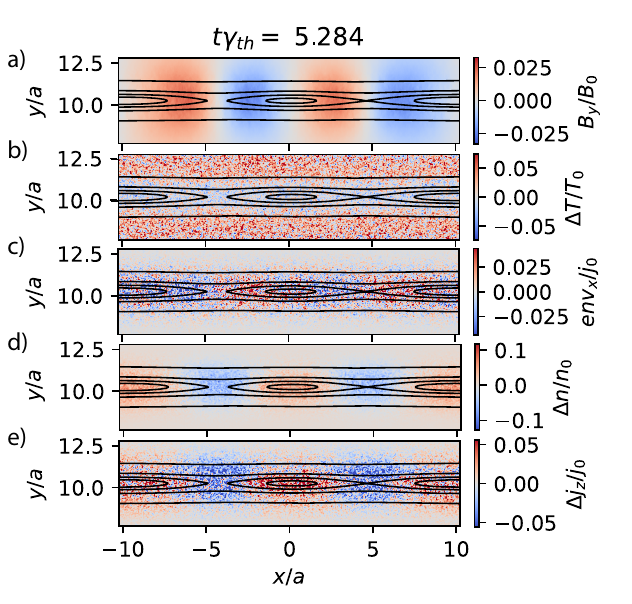}
  \caption{Map of the change in $B_y$, $n$, $j_z$, $T$, and $nv_x$ as a function of space for the simulation with $T/m_e c^2 = 0.00125$ and $a/\rho_L = 2.5$ at a late enough time where the signals are visible, but the growth is still in the linear stage.  Selected contours of magnetic flux overlaid to highlight the magnetic islands.}
\label{fig:modes}
\end{figure}

To better understand how to characterize tearing instability (before it reaches a strongly nonlinear stage), we present in Figure \ref{fig:modes} a spatial map of several quantities that characterize the tearing mode, with selected contours of magnetic flux overlaid to highlight the magnetic islands. We have chosen $B_y$ to measure the growth rates because the signal is visible at times as early as $t\gamma_{th}=3$, however, after around $t\gamma_{th}=5$, one can see in Figure \ref{fig:classicgrowth} (b) that a majority of the energy is being transferred to the kinetic energy in the Harris sheet. This energy goes to both heating and bulk flows. We show a map of $B_y$ in Figure \ref{fig:modes} (a) similar to what we saw in Figure \ref{fig:classicby}, but at $t\gamma_{th}=5.284$ and zoomed in on the current sheet. The energy is mainly converted into thermal energy. The temperature is shown in Figure \ref{fig:modes} (b), where there is an overall heating with cooling at the x-points (eg. at $x/a\approx-5, y/a \approx 5$). The energy going into the bulk flows includes a flow in the $x$ direction away from the x-points and toward the o-points (eg. at $x/a\approx -10, y/a \approx 0$), which can be seen in Figure \ref{fig:modes} (c). As the plasma moves with this flow, the density decreases at the x-points, and increases at the o-points as seen in Figure \ref{fig:modes} (d). This flow also drags the out-of-plane current with it as seen in Figure \ref{fig:modes} (e). The total kinetic energy in the current therefore also increases. 

Although we do not plot this here, the energy associated with the quantities in Figure \ref{fig:modes} (when applicable) all grow at the same growth rate during the linear stage, taking energy from the $B_x$ component of the magnetic field outside of the current sheets. We now report how much energy was transferred to each quantity by $t\gamma_{th}=5.284$, the time associated with Figure \ref{fig:modes}. The source of free energy is in the $B_x$ component; the total energy in $B_x$ drops by a factor of $3.3\times 10^{-4}$ its value at $t\gamma_{th}=0.48$. The energy predominantly goes to the thermal energy i.e. about $90\%$ plus an additional increase of $16\%$ due to numerical heating, while $13\%$ of the energy goes into $B_y$. The energy going into the bulk flows is about an order of magnitude less, about equally distributed between $1.3\%$ in the out-of-plane direction associated with the current, and $1.2\%$ in the in-plane directions associated with reconnection outflows along the $x$ direction. The energy going into $E_z$ is even less and the signal is not visible. This energy distribution between the different quantities is consistent with Figure \ref{fig:classicgrowth} which shows that the loss of energy in the total magnetic field in purple (predominantly associated with $B_x$) matches the gain in kinetic energy (predominantly thermal energy) \citep{Zenitani2017,Pucci2018b}. The energy in $B_y$ is about an order of magnitude less at $t\gamma_{th}=5.284$. At later times, all of these quantities convert the linear wave number mode to lower wave number modes and eventually evolve into a fast-growing nonlinear state of multiple interacting modes. 

\begin{figure}
\includegraphics[width=\textwidth]  {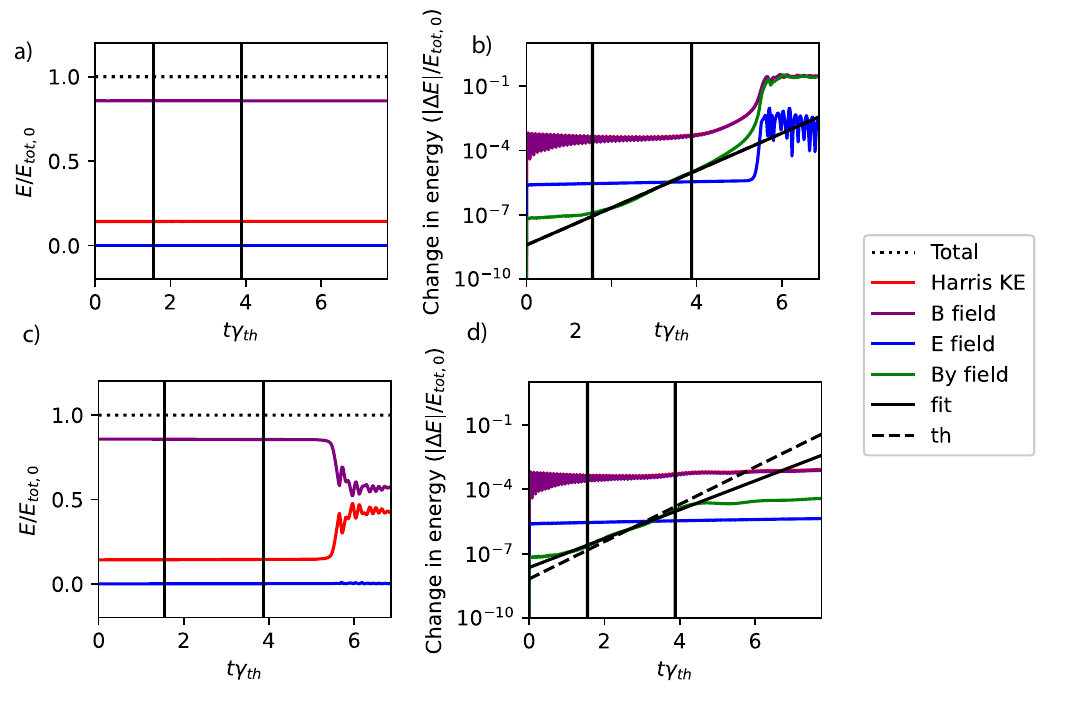}
  \caption{Evolution of the energy for the simulation with $T/m_e c^2 = 0.005$ and $a/\rho_{L,C} = 5$ (for $L_x = L_y/2$ above, where the growth saturates early, and for $L_x = L_y$ below, where it does not) in the Harris sheet electrons/positrons, electric, and magnetic fields, as well as the $y$ component of the magnetic field that characterizes the tearing growth rate. A fit of growth is plotted in solid black and the theoretical growth rate given by \eq{classicaltear2} in the dashed line.}
\label{fig:saturatedgrowth}
\end{figure}

Unlike the $a/\rho_{L,C} = 2.5$ case, when $a/\rho_{L,C} = 5$ the instability saturates before significant energy is released, i.e., before the fast-growing nonlinear stage of the instability is reached. For example, we show in Figure \ref{fig:saturatedgrowth} (a-b) the energy evolution for the case with $T/m_ec^2 = 0.005$. We can measure a growth rate $\gamma_m a/c = 0.00150$ which matches theory $\gamma_{th} a/c = 0.00194$, in the linear stage (between $t \gamma_{th} = 1.55-3.88$). However, after $t\gamma_{th} \sim 4.23$ the growth saturates.  The evolution continues without significant growth up to $t\gamma_{th} \sim 6.87$. We would like to point out that this saturation effect is dependent on the noise. We performed a similar set of simulations, not presented here, with much fewer particles-per-cell that were noisier and less accurate but obtained the same growth rates. In this noisier case, the signal was able to grow to the fast-growing nonlinear stage without saturating. 

We found however that by increasing the length of the box to $L_x = L_y$ (twice as long), we get a similar linear growth rate $\gamma_m a/c = 0.00194$ (matching theory almost perfectly). The wave number also remains consistent with theory with $ka = 2\pi m a/(2 L_x) \approx 0.6$ (now with a higher $m=4$). However, in this case, the instability does reach a fast-growing nonlinear stage. As we saw previously, the growth rate increases until it reaches $\gamma a/c = 0.0296$ close to the prediction corresponding to $a/\rho_{L,C} = 1$, $\gamma a/c = 0.0217$.  In Figure \ref{fig:saturatedgrowth}, we highlight the difference between the case with the smaller box ($L_x=L_y/2$) in Figure \ref{fig:saturatedgrowth}(a-b) and an identical case 
 except ($L_x=L_y$) in Figure \ref{fig:saturatedgrowth}(c-d). In the case with the larger box, significant energy is released as shown in Figure \ref{fig:saturatedgrowth}(c), and the nonlinear growth rate is measurable in Figure \ref{fig:saturatedgrowth}(d).

 \begin{figure}
  \centering
  \includegraphics[width=\textwidth]
  {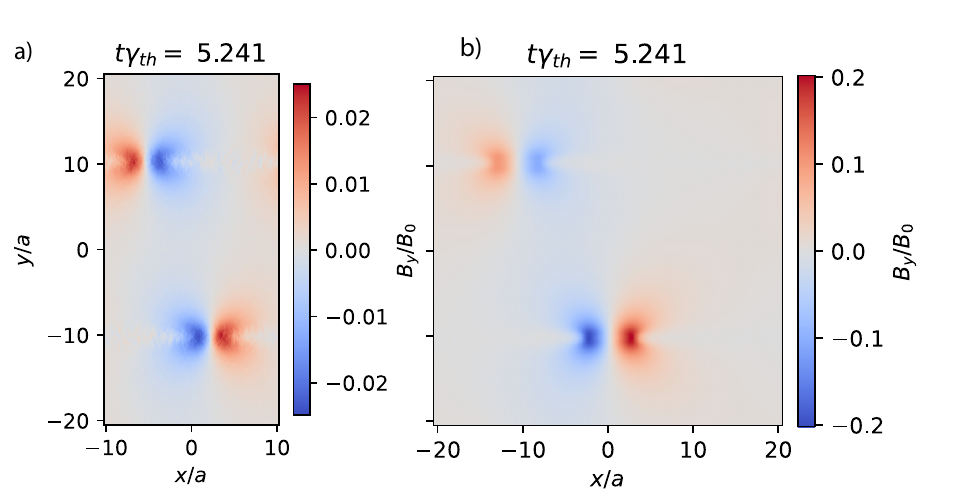}
  \caption{Map of $B_y$ as a function of space for the simulation with $T/m_e c^2 = 0.005$ and $a/\rho_{L,C} = 5$ during saturation 
  $L_x = L_y/2$ (left), and while transitioning into the fast-growing nonlinear stage $L_x = L_y$ (right).}
\label{fig:saturationmap}
\end{figure}

We illustrate in Figure \ref{fig:saturationmap} the time right before the nonlinear phase (at  $t\gamma_{th}=5.241$), where either the growth of $B_y$ saturates (when $L_x=L_y/2$) (a) or it blows up (when $L_x=L_y$) (b). In both cases, the tearing has moved from the linear $ka \approx 1/\sqrt{3}$ to the lowest $k$ that fits in the box (only 1 magnetic island). Furthermore, we note that at this stage, the current sheets begin to interact. Once again we see that the nonlinear stage matches the prediction from \cite{Hoshino2021}. Assuming $ka=1/\sqrt{3}$, we calculate the normalized wavelength $k\rho_{L,C} \equiv 0.12$. While the $B_y/B_0$ in Figure \ref{fig:saturationmap}(a) never exceeds this value and thus no explosive reconnection phase is observed, it exceeds this value in Figure \ref{fig:saturationmap}(b) and we do see an explosive phase.

From the start of the simulation, the tension from the bent magnetic field lines pulls the plasma toward the center of the magnetic islands, driving the instability. Meanwhile, the upstream magnetic field is also bent providing a stabilizing force on the inflow. During the linear phase of the tearing instability, the driving force is stronger than the stabilizing force. However, in the nonlinear regime, the stabilizing force can dominate. In the simulation with the large box, the aspect ratio of the island which is proportional to $L_x/a$ is also larger, and thus the driving force which is proportional to $1/a$ remains large compared to the stabilizing force which is proportional to $1/L_x$. This argument for saturation may also explain the transfer we see from the Zelenyi prediction $ka = 1/\sqrt{3}$ to the smallest mode that fits in the box $ka = 2\pi a/L_x$.

We find similar results for all of our simulations. For all temperatures with $a/\rho_{L,C} = 2.5$, we measure the growth rates in the range $t \omega_{pe} = 3.30-4.39 \omega_{pe}/\gamma_{th}$. We find that the measured growth rate $\gamma_m$ matches the theory $\gamma_{th}$ shown in Table \ref{table}. We also measure the growth rates for the non-linear stage $\gamma_{m,nl}$,
in the ranges $t_{st,nl}-t_{fi,nl}$ also found in Table \ref{table}.  We find that the nonlinear growth rate matches the linear prediction for $a/\rho_{L,C} = 1$.
For the cases with $a/\rho_{L,C} = 5$ we measure the growth rate that is consistent with the theory (both in Table \ref{table}) in the range $t \omega_{pe} = 1.55-3.88 \omega_{pe}/\gamma_{th}$, which is earlier than the interval used in the set of simulations with $a/\rho_{L,C} = 2.55$ because we have less noise due to the increased particles-per-cell. The peak growth rate for the nonlinear stage is again measured for the simulations with $L_x=L_y$ and is reported along with the time range of the measurement in Table \ref{table}. The growth rates in the linear stage of these simulations also match the theory well and are listed in Table \ref{table}. 

\begin{figure}
  \centering
  \includegraphics[width=\textwidth]{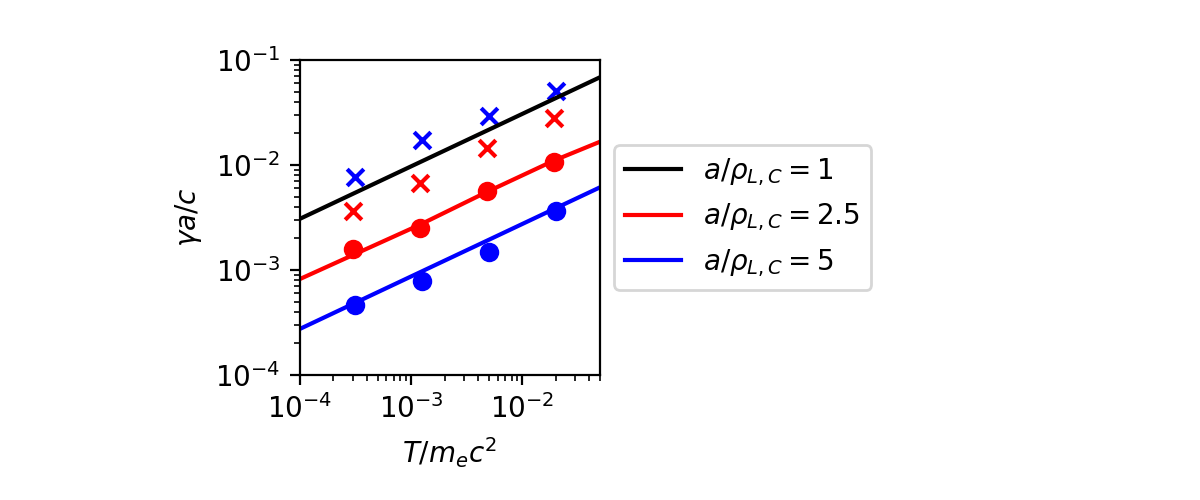}
  \caption{
  Measurement of the tearing growth rate in the linear  (solid circles) and non-linear (x's) stages, along with prediction from \eq{classicaltear2}, for $a/\rho_{L,C}=2.5$ (red), and $5$ (blue). The solid black line is the prediction for $a/\rho_{L,C}=1$.  All blue markers correspond to simulations with $a/\rho_{L,C}=5$ but the symbols correspond to different simulations; blue circles correspond to simulations with $L_x = L_y/2$, while blue x's correspond to simulations with $L_x = L_y$ where a nonlinear growth rate can be measured.
  }
\label{fig:classical}
\end{figure}

These results are summarized in Figure \ref{fig:classical}. We indicate the growth rates for the linear stage with solid circles $a/\rho_{L,C} = 2.5$ (red) and $a/\rho_{L,C} = 5$ (blue), and the corresponding theory \eq{classicaltear2} is plotted as a line with the same color. These measurements match remarkably well with the theory. We indicate the growth rates for the nonlinear stage with x's for $a/\rho_{L,C} = 2.5$ and $a/\rho_{L,C} = 5$ (Measured from simulation with $L_x=L_y$), with the same color scheme. The nonlinear growth rate can be well estimated by the black line which corresponds to \eq{classicaltear2} with $a/\rho_{L,C} = 1$.

The fast-growing nonlinear growth rate when $a/\rho_{L,C}=5$ is a factor close to $5^{3/2} \approx 11$ faster than the linear growth rate. However, earlier in this section, we hypothesized that although a background plasma would not affect the linear growth rate, the fast-growing nonlinear growth rate can be limited by the background. The linear growth rate is a function of $\rho_{L,C}/a \approx d_{e,C}/a$ [\eq{classicaltear2}]. While the fast-growing nonlinear growth rate can be estimated by the linear prediction assuming $\rho_{L,C}/a=1$, with a background, we predict it is given by the linear prediction replacing $\rho_{L,C}/a$ with the background inertial length $d_{e,C}(n_b)/a = \rho_{L,C}/a \sqrt{n_0/n_b}$, rather than $1$.
To test this hypothesis, we performed a simulation identical to the case with $a/\rho_{L,C}=5$, $T/m_ec^2 = 0.005$, and $L_x = L_y$, but with a background density of $n_b/n_0 = 0.1$.
The simulation showed both a similar linear growth rate and a slower nonlinear growth rate, that match this prediction. Using a low pass filter described in the next section to mitigate noise from the background population, we measure a linear growth rate of $\gamma_m a/c = 0.00197$, which is consistent with the theoretical value of $\gamma_{th} a/c = 0.00194$. The fast-growing nonlinear growth rate was measured as $\gamma_{m,nl} a/c = 0.0297$ for the case with no background, close to the linear prediction for $a/\rho_L = 1$, namely $\gamma a/c = 0.0217$.
With the background density $n_b/n_0 = 0.1$, the ratio $a/d_{e,C}(n_b) = 5\sqrt{0.1}$, suggesting a nonlinear growth rate $\gamma_{nl} a/c = 0.0109$, about half the growth rate for the case with no background. The measured nonlinear growth rate was $\gamma_{m,nl} a/c = 0.0095$, matching our predictions.

\subsection{Relativistic tearing}

\begin{table}
  \begin{center}
\def~{\hphantom{0}}
  \begin{tabular}{lcccccccccc}
      $a/\rho_{L,C}$  & $T/m_e c^2$    &   $u_d/c$  & 
      $\gamma_{th} a/c$ & $t_{\rm{st}}\gamma_{th}$ & $t_{\rm{fi}}\gamma_{th}$   & $\gamma_m a/c$ & $L_x/L_y$&
      $t_{\rm{st,nl}}\gamma_{th}$ & $t_{\rm{fi,nl}}\gamma_{th}$  & $\gamma_{m,nl} a/c$\\[3pt]
       ~0.148~   & $2.75 \times 10^{-5}$ & 0.05 & 0.00161~ & 2.71& 5.43& 0.00287~&1/2&-&-&-\\
       ~0.297~   & $1.10 \times 10^{-4}$ & 0.05 & 0.00322~& 2.71& 5.43 & 0.00438~&1/2&-&-&-\\
       ~0.593~   & $4.39 \times 10^{-4}$ & 0.05 & 0.00644~& 2.71& 5.43 & 0.00659~&1/2&-&-&-\\
       ~1.19~~   & $1.76 \times 10^{-3}$  & 0.05 & 0.00997~& 5.46& 6.30 & 0.00791~&1/2&10.59&10.92&0.0105\\
       ~2.37~~   & $7.03 \times 10^{-3}$ & 0.05 & 0.00705~& 2.97& 5.94& 0.00670~&1/2&~7.43&~7.55&0.0183\\
       ~4.74~~   & $2.81 \times 10^{-2}$ & 0.05 & 0.00499~& 2.10& 4.20 & 0.00550~&1/2&-&-&-\\
       ~9.49~~   & 0.113~~~~~ & 0.05 & 0.00353~& 5.48& 5.52& 0.00275~&1/2&-&-&-\\
       ~0.0371   & $2.75 \times 10^{-5}$ & 0.2~ & 0.00161~& 2.15& 4.30 & 0.00237~&1/2&-&-&-\\
       ~0.0741   & $1.10 \times 10^{-4}$ & 0.2~ & 0.00322~& 2.15& 4.30 & 0.00473~&1/2&-&-&-\\
       ~0.148~   & $4.39 \times 10^{-4}$ & 0.2~ & 0.00644~& 2.15& 4.30 & 0.00927~&1/2&-&-&-\\
       ~0.297~   & $1.76 \times 10^{-3}$ & 0.2~ & 0.0129~~& 2.15& 4.30 & 0.01683~&1/2&-&-&-\\
       ~0.593~   & $7.03 \times 10^{-3}$ & 0.2~ & 0.0258~~& 2.15& 4.30 & 0.02449~&1/2&-&-&-\\
       ~1.19~   & $2.81 \times 10^{-2}$ & 0.2~ & 0.0399~~& 2.94& 5.89 & 0.03240~&1/2&10.66&10.99&0.0395\\
       ~2.37~~   & 0.113~~~~~ & 0.2~ & 0.0282~~& 5.89& 6.36 & 0.03378~&1/2&~6.34&~6.59&0.0753\\
       ~4.74~~   & 0.45~~~~~~ & 0.2~ & 0.0295~~& 6.16& 7.39 & 0.0220~~~&1/2&~8.32&~8.50&0.0770\\
       ~9.49~~   & 1.8~~~~~~~ & 0.2~ & 0.00295~& -& - & - &1/2&12.32&12.63&0.0848\\
       19.0~~~   & 7.2~~~~~~~ & 0.2~ & 0.00295~& -& - & - &1/2&17.87&18.17&0.0879\\
       ~0.0093   & $2.75 \times 10^{-5}$ & 0.8~ & 0.00161~& 3.84& 7.68 & 0.00191~&1/2&-&-&-\\
       ~0.0186   & $1.10 \times 10^{-4}$ & 0.8~ & 0.00322~& 3.84& 7.68 & 0.00383~&1/2&-&-&-\\
       ~0.0371   & $4.39 \times 10^{-4}$ & 0.8~ & 0.00644~& 3.84& 7.68 & 0.00766~&1/2&-&-&-\\
       ~0.0741   & $1.76 \times 10^{-3}$ & 0.8~ & 0.0129~~& 3.84& 7.68 & 0.0153~~&1/2&-&-&-\\
       ~0.148~   & $7.03 \times 10^{-3}$ & 0.8~ & 0.0258~~& 3.84& 7.68 & 0.0300~~&1/2&-&-&-\\
       ~0.297~   & $2.81 \times 10^{-2}$ & 0.8~ & 0.0515~~& 3.84& 7.68 & 0.0540~~&1/2&-&-&-\\
       ~0.593~   & 0.113~~~~~ & 0.8~ & 0.1030~~& 3.84& 7.68 & 0.0832~~&1/2&-&-&-\\
       ~1.19~~   & 0.45~~~~~~ & 0.8~ & 0.1336~~& 4.98& 9.96 & 0.0872~~&1/2&-&-&-\\
       ~2.37~~   & 1.8~~~~~~~ & 0.8~ & 0.1336~~& 4.98& 7.47 & 0.0863~~&1/2&-&-&-\\
       ~4.74~~   & 7.2~~~~~~~ & 0.8~ & 0.1336~~& 4.98& 9.96 & 0.0859~~&1/2&-&-&-\\
       ~2.37~~   & $7.03 \times 10^{-3}$ & 0.05 & 0.00705~& 2.97& 5.94 & 0.00711~&1&~6.48&~6.65&0.0297\\
       ~4.74~~   & $2.81 \times 10^{-2}$ & 0.05 & 0.00499~& 1.05& 4.20 & 0.00620~&1&~4.73&~4.78&0.0587\\
       ~9.49~~   & 0.113~~~~~ & 0.05 & 0.00353~& 1.11& 2.97 & 0.00382~&1&~7.33&~7.37&0.0760\\
       19.0~~~   & 0.45~~~~~~ & 0.05 & 0.00386~& 0.61& 2.24 & 0.00346~&1&12.04&12.06&0.0934\\
       37.9~~~   & 1.8~~~~~~~ & 0.05 & 0.00386~& 1.02& 2.54 & 0.00350~&1&~7.66&~7.69&0.0981\\
       75.9~~~   & 7.2~~~~~~~ & 0.05 & 0.00386~& 0.51& 3.05 & 0.00402~&1&~7.45&~7.46&0.0989\\
       ~2.37~~   & 0.113~~~~~ & 0.2~ & 0.0282~~& 2.35& 5.89 & 0.0366~~&1&~5.42&~5.77&0.0718\\
       ~4.74~~   & 0.45~~~~~~ & 0.2~ & 0.0295~~& 1.85& 3.70 & 0.0277~~&1&~7.39&~7.70&0.0961\\
       ~9.49~~   & 1.8~~~~~~~ & 0.2~ & 0.0295~~& 3.08& 6.16 & 0.0206~~ &1&~8.32&~8.93&0.0946\\
       19.0~~~   & 7.2~~~~~~~ & 0.2~ & 0.0295~~& 3.08& 6.16 & 0.0200~~ &1&~7.39&~7.55&0.1031\\
       ~9.49~~   & 0.113~~~~~ & 0.05 & 0.00353~& 0.37& 1.49 & 0.00377~&2&~1.88&~1.91&0.0834\\
       19.0~~~   & 0.45~~~~~~ & 0.05 & 0.00386~& 0.61& 2.03 & 0.00272~ &2&~5.33&~5.39&0.0865\\
       37.9~~~   & 1.8~~~~~~~ & 0.05 & 0.00386~& 1.02& 3.05 & 0.00278~ &2&~5.17&~5.19&0.1016\\
       75.9~~~   & 7.2~~~~~~~ & 0.05 & 0.00386~& 1.02& 3.05 & 0.00308~ &2&~5.37&~5.39&0.0956\\
  \end{tabular}
  \caption{Parameters for the relativistic set of simulations including the theoretical linear growth rate $\gamma_{th}$ given by \eq{classicaltear2} with $\rho_{L,C}/a=1$ when $\rho_{L,C}/a < 1$, given by \eq{classicaltear} when $\rho_{L,C}/a > 1$ and $T/m_e c^2 < 0.15$, and given by \eq{reltear} when $T/m_e c^2 > 0.15$. The linear growth rate $\gamma_m$ all for the standard simulations with $L_x = L_y/2$ is measured between the start time $t_{\rm{st}}$ and the finish time $t_{\rm{fi}}$, and for simulations with $L_x = L_y$ or $2L_y$ a linear growth rate measured after doing a low pass filter over the same time interval. The fast-growing nonlinear growth rate $\gamma_{m,nl}$ is measured between $t_{\rm{st,nl}}$ and $t_{\rm{fi,nl}}$.}
  \label{table2}
  \end{center}
\end{table}
In this section, we examine a wider range of temperatures while keeping $u_d/c$ constant (instead of $\rho_{L,C}$ like the previous section); from $T/m_e c^2 = 2.74\times 10^{-5} \ll 1$ to $T/m_e c^2 = 7.2 \gg 1$, separated by factors of 4, thus exploring a range of both classical and relativistic temperatures. 
We examine three cases with $u_d/c=0.8, 0.2,$ and $0.05$, which respectively correspond to $a/\rho_{L,R} = c/u_d  = 1.25, 5$ and $20$ for relativistic temperatures
(see Table \ref{table2} for a list of all the simulations.).
Here we explore regimes beyond the scope of the Zelenyi model, i.e. \eqs{classrho}{relrho}.
When we keep $\rho_{L,R}/a$ constant while varying $T/m_ec^2$, as a consequence we are also varying $\rho_{L,C}/a$, and for increasingly small temperatures, $a/\rho_{L,C}$ decreases. Therefore, for many of our simulations $a/\rho_{L,C}$ is smaller than $1$, and since the temperature is classical, the assumption that $\rho_L/a \ll 1$ breaks down.

Note that it is, in fact, possible to have a current sheet where $a < \rho_{L,C}$. When there are strong gradients in the magnetic field and few particles to sustain a current, the particles can be accelerated beyond the thermal velocity. The current is thus composed of beams of particles trapped in a magnetic potential without the possibility of making Larmor orbits.

For the $u_d/c = 0.8$ case, $a/\rho_{L,R} = 1.25$, so even for large temperatures the assumption $\rho_L/a \ll 1$ breaks down. 
However, this region is in the scope of the predictions from \cite{Hoshino2020} included in \eq{relrho}. This model predicts that $\rho_{L,R}/a = u_d/c \approx 0.8$ is the optimal value for the maximum growth rate (higher $u_d/c$ leads to a suppression of the instability.).

For each simulation, we use 1024 particles-per-cell, $L_y/a = 21.4$, and $L_x/a= 10.7$ with a resolution of 18 grid cells per $a$. We always choose a time step to satisfy the Courant condition. Let us first consider the classical regime where $T/m_ec^2 \ll 1$. Just as in the previous section, we calculate growth rates, both in the linear stage and in the nonlinear stage where the growth rate rapidly increases, by calculating a line of best fit of the $B_y$ component of the energy. Note that no faster nonlinear stage is found for cases where $\rho_{L,C}/a > 1$ and the assumption that $\rho_L/a \ll 1$ breaks down. This is expected because in these cases, the growth rate is already at its maximum with respect to $\rho_{L,C}/a$. For some cases, we also perform identical simulations with increased length in the $x$ direction $L_x = L_y$ and $2L_y$, which will be identified in the text when they are used.

Before discussing the measurement of the growth rate in the relativistic regime, let us briefly discuss the expected particle noise that seeds the instability. In a classical Maxwellian distribution, there are fewer particles with high $v/c$ leading to predominantly low $k$ noise in the magnetic field. This is due to the large inter-spacial distance between these energetic particles. In contrast, for ultra-relativistic temperatures, there is only weak $k$ dependence as nearly all particles have the same value of $v/c \approx 1$. Therefore, in the relativistic regime, the particle noise with high $ka$ contributes significantly to the magnetic energy in $B_y^2/4\pi$, making it difficult to measure the growth rate of the signal simply from the evolution of the energy in $B_y^2/4\pi$. While one might expect a similar effect when $u_d/c$, rather than $T/m_ec^2$, becomes relativistic, the faster growth rates associated with faster $u_d/c$ can more easily overcome the noise. Furthermore, the in-plane thermal noise is reduced by $1/\Gamma_d$ after being boosted into the moving frame. An increasing temperature, on the other hand, keeping $u_d/c$ constant, can only slow the growth rate.

\begin{figure}
  \includegraphics[width=\textwidth]  {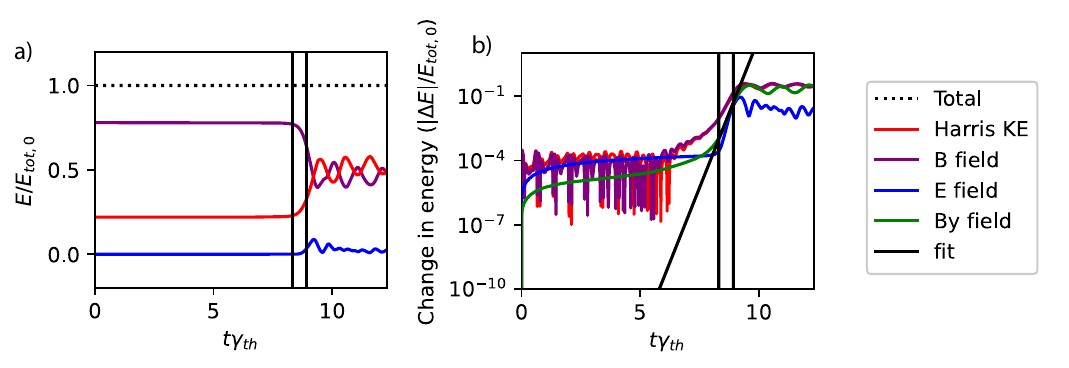}
  \caption{Evolution of the energy for the simulation with $T/m_e c^2 = 2$ and $u_d/c = 0.2$ (with $L_x = L_y$) in the Harris sheet electrons/positrons, electric, and magnetic fields, as well as the $y$ component of the magnetic field that characterizes the tearing growth rate. No linear growth rate is measurable, but a fit of the fast-growing nonlinear growth is plotted in solid black.}
\label{fig:relativisticgrowth}
\end{figure}

An example of the evolution of the energy of the particles, electromagnetic fields, and the energy in the $B_y$ component of the magnetic field is shown in Figure \ref{fig:relativisticgrowth} for the case where $T/m_e c^2 = 2$ and $u_d/c = 0.2$. Note that in this case, we have doubled the length to $L_x = L_y$, which gives similar results to the standard case where $L_x = L_y/2$, but will help us to measure the growth rate. 
Like in the previous section, we normalize the time to the theoretical growth rate $\gamma_{th} = 0.0295$. However, here we have calculated the theoretical growth rate using the relativistic model, \eq{reltear}. As expected the noise of the magnetic field dominates throughout the linear stage, and a measurement cannot be taken.
Furthermore, the scale separation between the noise in the different energy channels is no longer significant. The energy in the $B_y$ component of the magnetic field is a factor of $v_T/c$ less than the electric field energy, which itself is a factor of $v_T/c$ less than total magnetic field and kinetic energy, as seen in Figure \ref{fig:classicgrowth} in the nonrelativistic case when $v_T\ll c$. With relativistic temperatures ($v_T \sim c$), this separation is no longer present, making a measurement more difficult.
However, around $t\gamma_{th} \sim 8$ the system reaches the nonlinear stage, and like in the classical case the growth rate increases rapidly and overcomes the noise. We can thus measure a nonlinear growth rate between $t\gamma_{th} \sim 8.3-8.9$ of $\gamma a/c = 0.0848$. Again, like in the classical case, Figure \ref{fig:relativisticgrowth} (a) shows in the nonlinear regime significant energy originally from the $B_x$ component of the electromagnetic field is converted to kinetic energy. 

\begin{figure}
  \centering
  \includegraphics[width=\textwidth]
  {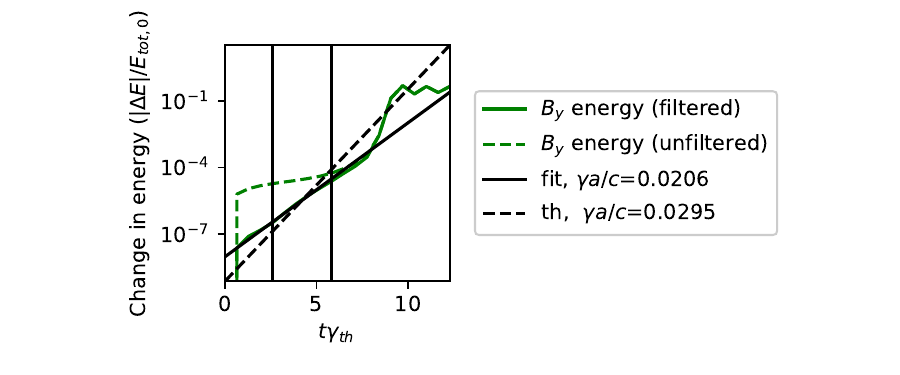}
  \caption{Evolution of the $B_y$ energy for the simulation with $T/m_e c^2 = 2$ and $u_d/c = 0.2$ (with $L_x = L_y$) unfiltered (dashed lines) and after performing a low pass filter only allowing the modes $m=6$ and below (solid lines). A fit of growth is plotted in solid black and the theoretical growth rate given by \eq{reltear} in the dashed line.}
\label{fig:relativisticfiltergrowth}
\end{figure}

To measure the linear growth rate, we put the magnetic field grid through a low pass filter, keeping only $ka \lesssim 1$ ($m \equiv k_x L_x/\pi$ and $k_y L_y/\pi \leqq 6$) which are the modes that we expect to grow and constitute our signal. To have a better resolution in $k$-space, we use simulations with $L_x = L_y$, i.e. double the length of our fiducial runs. In Figure \ref{fig:relativisticfiltergrowth}, we show the evolution of this filtered magnetic energy evolution compared to the curve of the unfiltered magnetic field shown in dashed lines, and we can measure a growth rate between $t\gamma_{th} \sim 3.1-6.2$ of $\gamma_m a/c = 0.0206$, which is comparable to the theoretical value $\gamma_{th} a/c = 0.0295$.

\begin{figure}
  \includegraphics[width=\textwidth]  {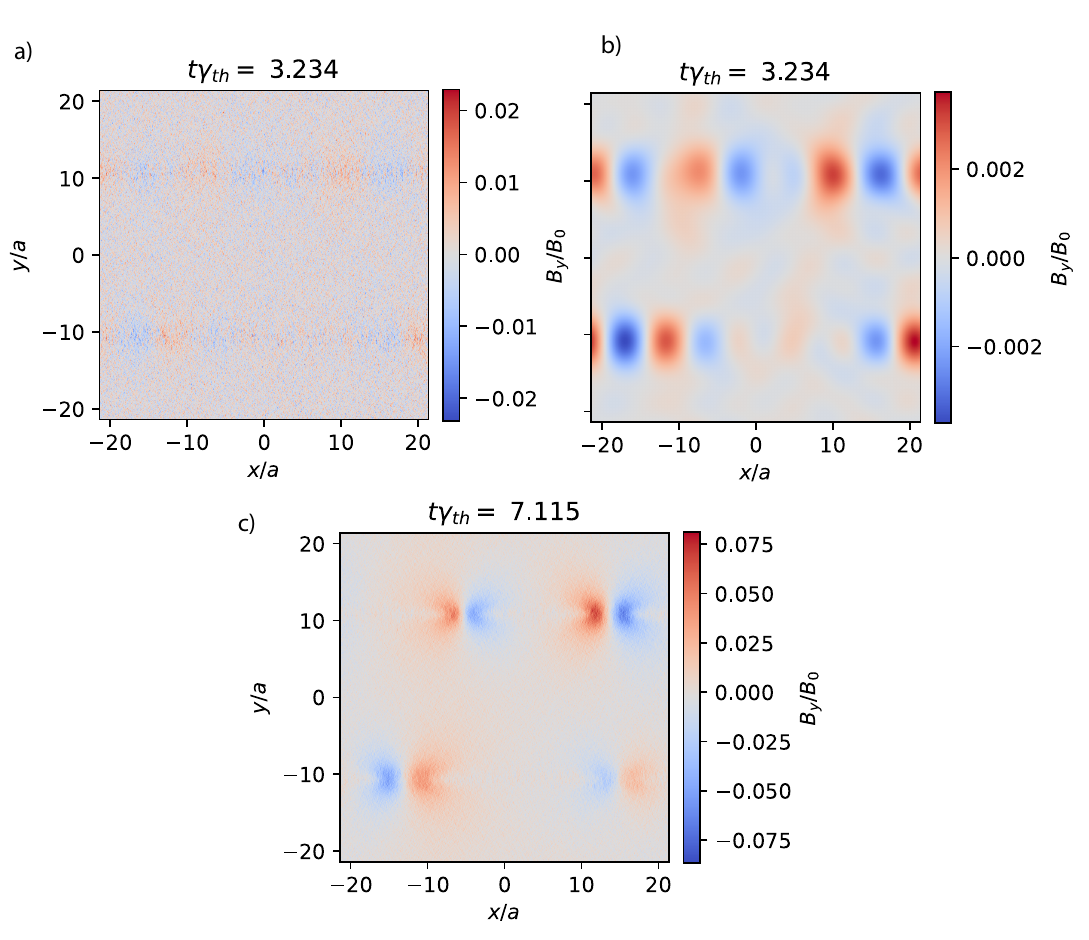} 
  \caption{Map of $B_y$ as a function of space for the simulation with $T/m_e c^2 = 2$ and $u_d/c = 0.2$ (with $L_x = L_y$) unfiltered (left) and after performing a low pass filter only allowing the modes $m=6$ and below (right), and at a later time where the smaller $ka$ modes begin to dominate.}
\label{fig:relativisticmap}
\end{figure}

Figure \ref{fig:relativisticmap} (a) illustrates the significant noise in the linear growth stage in a map of the magnetic field, while a low $k$ signal is visible. Figure \ref{fig:relativisticmap} (b) shows that the low pass filter removes this noise while retaining the low $k$ signal. Finally, Figure \ref{fig:relativisticmap} (c) shows the signal once it has grown beyond the noise. At this point, it has already reached a nonlinear stage, where the smallest $ka$ dominates and the growth rate begins to blow up.

\begin{figure}
  \centering
  \includegraphics[width=\textwidth]{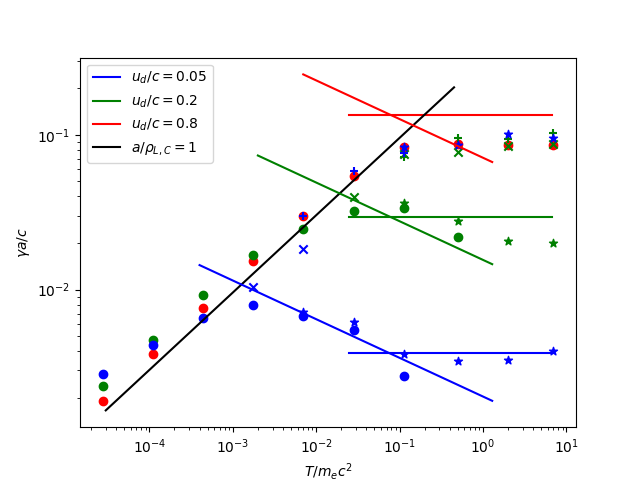}
  \caption{Measurement of the tearing growth rate in the linear (solid circles), and in non-linear (x's) stages, along with prediction in the classical \eq{classicaltear} (dashed lines) and relativistic \eq{reltear} (solid lines) temperature regimes for $u_d/c=0.05,0.2$, and $0.8$, along with the prediction for \eq{classicaltear2} when $a/\rho_{L,C}=1$ (solid black line). Stars represent simulations with $L_x=L_y$, where the growth rate was measured after performing a low pass filter. In addition to the points marked with x's, additional simulations measuring the non-linear growth with $L_x=L_y$ are marked with +'s, and $L_x=2L_y$ with stars.}
\label{fig:relativistic}
\end{figure}

In Figure \ref{fig:relativistic} we summarize the temperature dependence on the growth rate for the various values of $u_d/c$ from all our simulations, which can also be found in Table \ref{table2}. 
The standard measurements of the linear phase are marked by circles, while the simulations measured in larger boxes ($L_x=L_y$) using a low pass filter are marked by stars.
The nonlinear growth rate was also measured for the $u_d/c=0.2$ and $u_d=0.05$ cases, and the results are indicated by x's for the standard simulation with $L_x = L_y/2$, +'s for the simulations with $L_x = L_y$, and stars for $L_x = 2 L_y$.
This is an equivalent plot to Figure \ref{fig:classical} from the classical part of Section \ref{sec:results}, which is also the growth rate as a function of temperature. Figure \ref{fig:classical} would fit in the low temperature and $u_d/c$ regime (lower left corner of Figure \ref{fig:relativistic}), remembering that while here $a/\rho_{L,R} = u_d/c$ is held constant, in Figure \ref{fig:classical}, $a/\rho_{L,C}$ is held constant. 

Let us first examine the familiar classical regime of the $u_d/c = 0.05$ simulations for temperatures above $T/m_ec^2 \approx 2 \times 10^{-3}$ ($a/\rho_{L,C} =1.26$), where the simulated growth rates follow the prediction for the classical regime, \eq{classicaltear} (left blue line). 
For lower temperatures, the current thickness $a/\rho_{L,C} < 1$, and the growth rates fit the predictions for $\rho_{L,C}/a = 1$ (indicated by the black line). A better approximation, at least for electron-ion plasmas, is given by \cite{Pritchett1991}, who looks in the small $a/\rho_{L,C} \sim 1$ regime, finding a similar limit for $a/\rho_{L,C} \ll 1$. Also \cite{Brittnacher1995} finds an analytic expression that works well for both regimes of $a/\rho_{L,C}$.
Finally, for temperatures larger than $T/m_ec^2 \approx 0.45$, the growth rates follow \eq{reltear} (horizontal blue line).
Similarly, the $u_d/c = 0.2$ simulations follow \eq{classicaltear} between $T/m_ec^2 \approx 3 \times 10^{-2}$ ($a/\rho_{L,C} =1.22$) and $T/m_ec^2 \approx 0.45$ (left green line).
Note that the range of validity for \eq{classicaltear} is shorter than in the $u_d/c = 0.05$ case. For lower temperatures, the growth rates follow predictions for $\rho_{L,C}/a = 1$, and for relativistic temperatures beyond $T/m_ec^2 \approx 0.45$, they follow \eq{reltear} (horizontal green line). The growth rate does not match \eq{reltear} precisely but is smaller by a factor of $1.5$, a factor similar to the $\sim 1.7-2$ found in \cite{Hoshino2020} and \cite{Zenitani2007}, who only considered values of $u_d/c \ge 0.3$.
We can see in these curves that, as claimed in the previous section, for constant $u_d/c$, the growth rate reaches a peak near $a/\rho_{L,C} = 1$.

For $u_d/c = 0.8$ (points marked in red), at no point does the growth rate match \eq{classicaltear} (indicated by the red line on the left), as it is always true that $\rho_{L,C}/a > 1$, breaking the assumptions of the model. 
In the classical temperature regime, the growth rate matches the predicted value from \eq{classicaltear2} for $\rho_{L,C}/a = 1$ (black line), until $T/m_ec^2 \approx 0.1$ when the growth rate becomes independent of the temperature as predicted by \eq{reltear} (indicated by the horizontal red line) for relativistic plasmas. Like in the $u_d/c =0.2$ case, the prediction overestimates the growth rate by a factor of $\sim 1.5$. We also expect, as claimed in the previous section, that for constant $T/m_ec^2$, a peak growth rate occurs near $a/\rho_{L,C}=1$. In Hoshino's model for the relativistic temperature regime i.e. \eq{reltear}, the growth rate for small $u_d/c$ is proportional to $(u_d/c)^{3/2}$, and for large $u_d/c$ (implying $a/\rho_{L,R}<1$) it is proportional to $(u_d/c)^{-1} \sim 1/\Gamma_d$, leading to a peak in between at moderate $u_d/c \sim 1$. For the coldest temperatures simulated, when $a/\rho_{L,C}<1$ the growth rate also decreases with $1/u_d$, and thus a peak growth rate also exists when $a/\rho_{L,C} \sim 1$.

Although not shown in the figure, we performed one simulation identical to our previous simulations with $T/m_ec^2 = 2$ and $L_x = L_y$, but this time with $u_d/c = 10$ to confirm Hoshino's model for large drift velocities. We were able to measure a growth rate between $t\gamma_{th} = 2.8-8.4$ of $\gamma_m a/c = 0.027$ using both the growth with and without the low pass filter (The thermal noise is greatly reduced in the boosted frame.). This value is consistent with the theoretical value $\gamma_m a/c = 0.034$ from \eq{reltear}, with only a factor of 1.3 overestimation. The wave number at the start of the measurement was $ka = 0.47$, which is consistent with the wave numbers $ka = 0.3 - 0.5$ reported in \cite{Hoshino2020}.

We measure the nonlinear growth rate in simulations where $a/\rho_{L,C} > 1$, for all cases where $u_d/c = 0.2$ and several cases where $u_d/c = 0.05$.
In the cases when $u_d/c = 0.05$ and $T/m_ec^2 > 0.028$ ($a/\rho_{L,C} > 4.7$), the growth rate saturates early and therefore there is no measurement presented. Like in the previous section, we double the length, increasing to $L_x = L_y$, which allows a wave number as low as $k a = 0.16$ at the $m=1$ mode, and find that the growth reaches the fast-growing nonlinear stage where significant energy is released before saturation occurs. Once again when $u_d/c = 0.05$ and $T/m_ec^2 > 0.5$ ($a/\rho_{L,C} > 20$), the growth rate saturates early for the $L_x = L_y$ case, and likewise we double the length again ($L_x = 2 L_y$), which allows a wave number as low as $k a = 0.08$ at the $m=1$ mode and reach the fast-growing nonlinear stage. (The growth rate does eventually reach the fast-growing nonlinear stage without doubling $L_x$ due to a slow growth after saturation.) Beyond this temperature is considered the relativistic regime, and $a/\rho_{L,R} = 20$. Unsurprisingly, we do not see any more early saturation as we increase the temperature. We expect the fast-growing nonlinear stage can always be reached, for a sufficiently long system. An important question remains; how does this critical length scale with the $\rho_L/a$?

At the nonlinear stage, the growth rate increases to the prediction from \eq{classicaltear2} for $\rho_{L,C}/a = 1$ (black line) in the nonrelativistic regime ($T/m_ec^2 < 0.1$), similar to the observations from the previous section. In the relativistic regime, the growth rate increases to the same value as the linear growth rate measured in the $u_d/c = 0.8$ case. As $u_d/c = \rho_{L,R}/a$, this is equivalent to the prediction of $\rho_{L,R}/a \sim 1$ for the relativistic regime, and thus one can make a generalized statement; in the nonlinear stage, the growth rate rises until it reaches the prediction for $\rho_L/a \sim 1$.

\section{Astrophysical limits on tearing}\label{sec:astrolim}
For various astrophysical environments, one can put a limit on the thinnest steady state current sheet that can form before tearing grows and disrupts the current sheet, using the prediction for the tearing growth rate. This limit is predicated on the assumption that, for a system with a size $L$, current formation occurs at a time scale slower than $\tau_F \sim L/v_A$, where $v_A$ is the Alfv\'en speed.
In our setup, based on pressure balance, $v_A = v_T$, and we will take the classical and ultrarelativistic limits,  $v_T = \sqrt{2T/m_e}$ and $v_T = c$ respectively.
The tearing instability grows faster as the thickness of the current sheet $a$ shrinks. If it reduces to a thickness $a$ where the growth rate reaches $\gamma \tau_F \sim 1$, the instability will occur before the current sheet can get any thinner.
We can thus calculate a minimum $a$ using \eq{classicaltear2} or the relativistic version \eq{reltear}, with  $\gamma \tau_F = 1$ at the fastest growing mode $ka \approx 1/\sqrt{3}$. We have found that the fastest-growing mode in our simulations remains at this value for large $L/a$ and make the assumption that this trend continues for increasing $L/a$.

Note that this limit follows the same assumptions of this study, a pair-plasma Harris current sheet with no guide field and the constant-$\psi$ approximation. When there is no guide field, a thin current sheet would likely be subject to the drift kink instability \citep{Zenitani2008}. Furthermore, a guide field is often present in instances of reconnection, but this only reduces the growth rate. We are thus making an upper limit on the minimum thickness of a current sheet. One should also take this as an order-of-magnitude estimate. An electron-ion plasma with a mass ratio would have a similar, but not equal growth rate as a pair plasma. Furthermore, we assume that the tearing instability is spontaneous rather than driven; the growth rate can be enhanced due to the injection of Poynting flux.

We thus find the minimum $a$ for the classical regime
\begin{equation}
\label{classicalalim}
        \frac{a_{min}}{L} \approx C_C  \left(\frac{\Omega_c L}{c}\right)^{-3/5} \left(\frac{T}{m_e c^2}\right)^{3/10}\left(\frac{L}{v_A \tau_F}\right)^{-2/5},
\end{equation}
and for the relativistic regime
\begin{equation}
\label{relalim}
        \frac{a_{min}}{L} \approx C_R  \left(\frac{\Omega_c L}{c}\right)^{-3/5} \left(\frac{T}{m_e c^2}\right)^{3/5} \left(\frac{L}{v_A \tau_F}\right)^{-2/5},
\end{equation}
where the respective constants are $C_C = 2^{7/10}3^{-3/5}\pi^{-1/5} \approx 0.67$ and $C_R = 2^{8/5}3^{-3/5}\pi^{-2/5} \approx 0.99$. As this is an order of magnitude estimate, and these constants are close to unity, we can neglect them. The normalized length scale
\begin{equation}
\label{normL}
    \frac{\Omega_c L}{c} \approx 1.81 \times 10^{15}\frac{B}{\text{1 Gauss}}\frac{L}{\text{1 Parsec}} 
\end{equation}
tends to be a large number, for many astrophysical contexts, and therefore the ratio $a/L$ is expected to be small.

\begin{figure}
  \centering
  \includegraphics[width=0.49\textwidth]{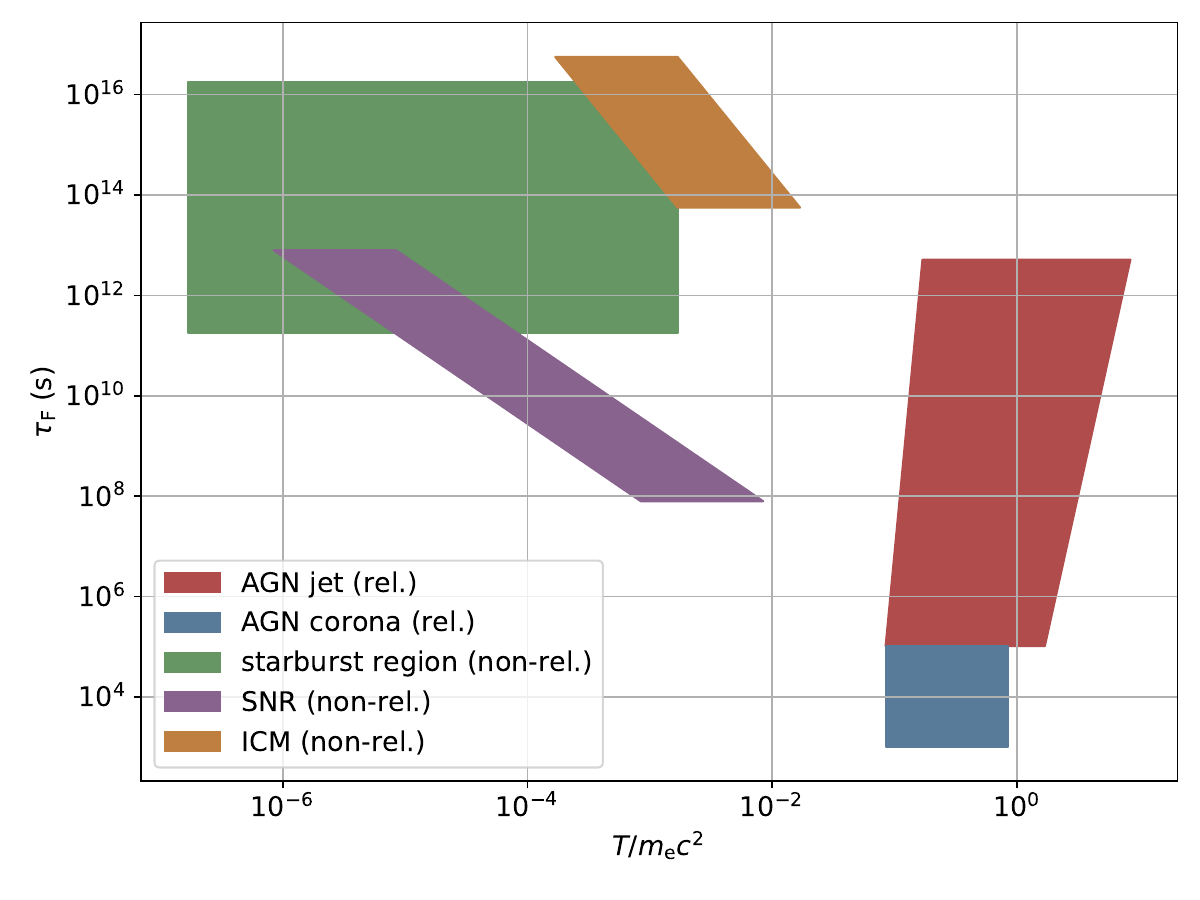}
  \includegraphics[width=0.49\textwidth]{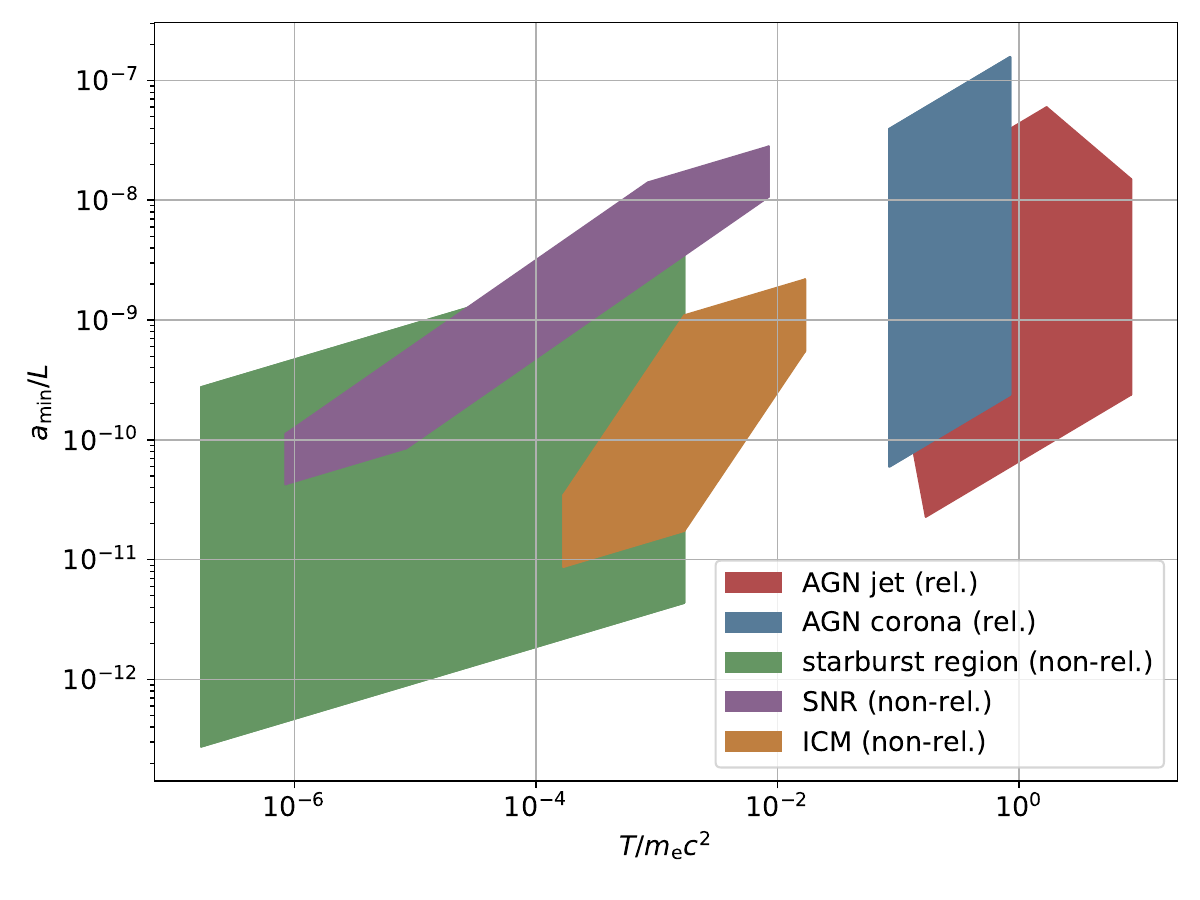}
  \caption{Potential ranges of the minimum thickness for a set of astrophysical environments based on the characteristic orders of magnitude of the system parameters $L$, $B$ and $T$. Dependent on the astrophysical environment [such as active galactic nucleus (AGN), supernova remnant (SNR) or intracluster medium (ICM)] \eq{classicalalim} and \eq{relalim} are adopted for relativistic and non-relativistic temperatures, respectively, as well as a minimal current formation time as given by $\tau_F = L/v_A$ with $v_A=v_T$ for the relativistic and non-relativistic limits.
  }
\label{fig:thickness}
\end{figure}


Figure \ref{fig:thickness} shows the current formation time scale $\tau_F$ and the predicted minimum current sheet thickness $a_{min}$ normalized to $L$ [using \eq{classicalalim} and \eq{relalim}] for various astrophysical regimes. A range of values are highlighted for each regime based on the typical orders of magnitude of the system size $L$, magnetic field strength $B$, and temperature $T$, assuming $\tau_F = L/v_A$. The ratio $a_{min}/L$ remains small for all of the regimes considered, and we find that this ratio tends to become smaller for cooler temperature regimes.  Furthermore, the formation time is significantly lower for more relativistic regimes.

As we previously stated, the current formation time tends to be proportional to $\tau_F \sim L/v_A$. However, the very thin current sheets predicted in Figure \ref{fig:thickness} can take considerably longer to form ($\tau_F \gg L/v_A$). We expect reconnection to be more prominent for the regions with the shortest time $\tau_F$ before $a$ reaches $a_{min}$ (and tearing onsets). We therefore expect significant reconnection, a source of energetic particles, in the more relativistic temperatures, where the minimum current thickness is wider. In this regard, the AGN corona is the most promising source candidate for particle acceleration by reconnection.

On the other hand, $a/\rho_L$ tends to be very large, as the particle's Larmor radius is typically multiple orders of magnitude smaller than the astrophysical system size.
One can write the previous equations in terms of $a/\rho_L$ for the classical regime
\begin{equation}
\label{classicalanlim}
        \frac{a_{min}}{\rho_{L,C}} \approx \frac{C_C}{\sqrt{2}}  \left(\frac{\Omega_c L}{c}\right)^{2/5} \left(\frac{T}{m_e c^2}\right)^{-1/5} \left(\frac{L}{v_A \tau_F}\right)^{-2/5},
\end{equation}
and for the relativistic regime
\begin{equation}
\label{relanlim}
        \frac{a_{min}}{\rho_{L,R}} \approx \frac{C_R}{2}  \left(\frac{\Omega_c L}{c}\right)^{2/5} \left(\frac{T}{m_e c^2}\right)^{-2/5} \left(\frac{L}{v_A \tau_F}\right)^{-2/5}.
\end{equation}
For example, the minimum thickness for AGN parameters would be around $a_{min}/\rho_{L,R} \sim 40000$. Again this is an order of magnitude estimate, and as the constants in front are close to unity, we can neglect them. 

For such thick current sheets (with respect to the particles' Larmor radius), we have to extrapolate from the much thinner current sheets that we tested numerically using the theoretical expressions. The three cases shown in Figure \ref{fig:relativistic}, $u_d/c = 0.8, 0.2,$ and $0.05$, correspond to only $a/\rho_{L,R} = 1.25,5,$ and $20$ respectively. While we have put a limit on the minimum thickness at the astronomical scales, it is unlikely that current sheets of such a high aspect ratio would occur. We also expect smaller scales to occur within the context of turbulence \citep{Comisso2018}, or the nonlinear evolution of the tearing instability, as shown in our simulations.

Also, for very thick current sheets, collisions can play a role. For collisional tearing the growth rate is $\gamma a/v_A \sim S^{-1/2}$, where the Lunquist number $S \equiv a v_A/\eta \sim (a/r_e) (v_T/c) (T/m_e c^2)^{3/2}$, $\eta$ is the resistivity, and $r_e$ is the classical electron radius. This scaling for $S$ is based on the Spitzer resistivity, which is independent of density, and since $v_A = v_T$ is only a function of $T$ and $a$.
Equating this growth rate with the collisionless growth rates [\eqs{classicaltear}{reltear}], we find that the transition from collisional to collisionless occurs when the temperature exceeds a certain value, expressed in terms of $L$ using \eqs{classicalanlim}{relanlim}, roughly:
\begin{equation}
\label{Ttransclass}
    \frac{T}{m_e c^2} \gtrsim 0.1 \left(\frac{B}{\text{1 Gauss}}\right)^{18/29}\left(\frac{L}{\text{1 Parsec}}\right)^{8/29} \left(\frac{L}{v_A \tau_F}\right)^{-8/29},
\end{equation}
for the classical regime, or
\begin{equation}
\label{Ttransclass}
    \frac{T}{m_e c^2} \gtrsim 0.1 \left(\frac{B}{\text{1 Gauss}}\right)^{9/19}\left(\frac{L}{\text{1 Parsec}}\right)^{4/19} \left(\frac{L}{v_A \tau_F}\right)^{-4/19},
\end{equation}
for the relativistic regime. Therefore, for cold plasmas particularly in denser regions with strong magnetic fields such as e.g.\ starburst regions, collisional effects may determine the minimum current thickness $a_{min}$. On the other hand, these collisional effects can clearly be ruled out for the different high-temperature locations within an AGN.

\section{Conclusion}\label{sec:conclusion}
We have investigated the tearing instability for a collisionless pair plasma, starting from a Harris equilibrium and no guide field or background population for a range of temperatures and drift velocities, from the classical regime where $T/m_e c^2= 3\times 10^{-5}$ and $u_d/c = 0.05$ to the relativistic regime where $T/m_e c^2= 7.2$ and $u_d/c = 0.8$. The growth rates match the predictions from \cite{Zelenyi1979} including modifications by \cite{Hoshino2020} for relativistic drift velocities quite well for all the valid regimes ($a/\rho_L \gg 1$), with a dominant mode at $ka \approx 1/\sqrt{3}$. The close agreement between theory and simulation results shows that $a/\rho_L > 1$ (as opposed to $a/\rho_L \gg 1$) is a sufficient condition.
Our measurement of the growth rate for relativistic temperatures is not as precise, and this coincides with arguably less strict agreement with the theory.

We have found that as the instability progresses, the dominant mode shifts from the Zeleyni prediction $ka = 1/\sqrt{3}$ toward the longest wavelength that fits in the simulation box, and the instability tends to saturate when $L_x$ is below a threshold that depends on $a/\rho_L$.
We also find that in the nonlinear stage of the instability, when $a/\rho_L>1$, the growth rate increases up to a maximum rate around the prediction for $a/\rho_L=1$. In the other regime with thin current sheets where $a/\rho_L<1$, the growth rate is already at its maximum and can be estimated by the prediction for $a/\rho_L=1$.
We find that this growth rate can be limited in the presence of a background density to the linear prediction for $a/\rho_L=a/d_{e,C}(n_b)\approx (a/\rho_{L,C}) \sqrt{n_b/n_0}$.

Moreover, we have obtained a prediction for a minimum current thickness $a_{min}/L$ that can be formed before tearing breaks up a current sheet. This prediction has been applied to different astrophysical systems showing that the minimum current sheet thickness is multiple orders of magnitude smaller than the system size $L$. Hence, these thin current sheets can clearly not be realized in starburst regions or the intracluster medium (ICM) since their formation takes about the age of the Universe or longer. But in some relativistic environments of an active galactic nucleus (AGN)---in particular the AGN corona---even these thin structures can in principle be realized, so that we expect the occurrence of reconnection providing energetic particles. Recent observations \citep{IceCube2022_ngc1068} by the IceCube detector indicate high energy neutrinos from a particular AGN called NGC 1068, that originate from its AGN corona as proposed by e.g.~\cite{Inoue+2020, Kheirandish+2021, Eichmann+2022}. To produce these neutrinos in the first place, high-energy cosmic ray protons are needed that could be generated via reconnection. A more detailed investigation, beyond the scope of this work, is still needed to clarify the actual acceleration processes in these astrophysical systems.

Despite these conclusions, we acknowledge several assumptions we have made that do not always hold. This implies other regimes that require further investigation. We have assumed a pair plasma, so the effect of different mass ratios remains to be explored. We have also assumed that there is neither a guide field nor a background population. Furthermore, all simulations were performed in 2D. Other instabilities (eg. drift kink instability \citep{Zenitani2008}) can occur in a 3D model.

Our simulations were done all with a mass ratio of $1$. In a system with an electron-proton-dominated plasma, we expect similar results, as predicted by Zelenyi for thick current sheets. We have done simulations not presented here where $\rho_{L,p} < a$, and the growth rate matches Zelenyi's prediction. We have not explored the intermediate regime where $\rho_{L,p} > a > \rho_{L,e}$. The fast-growing nonlinear mode would in principle pass through this intermediate regime. One may still ask; what implications does the Hall term have on the system for thick current sheets?

When a strong enough guide field $B_z$ is included [such that $B_z/B_0 > (\rho_L/a)^{1/2}$], predictions show slower growth rates that scale as $\gamma \sim (\rho_L/a)^{2}B_0/B_z$ instead of $\gamma \sim (\rho_L/a)^{3/2}$, and when $\rho_L \ll d_e$ they can be even slower with $\gamma \sim (d_e/a)^{3}$. Comparable differences should occur for force-free initial conditions instead of a Harris equilibrium. We suspect similar conclusions in these regimes, but the differences remain out of the scope of this paper. The typical current sheet configuration is not well known for relevant astrophysical systems, so these differences remain an important open question.

Zelenyi predicted how a background plasma would affect the tearing instability, and concluded that the background could be neglected for densities below a critical value $n_b/n_0 \sim (\gamma_{th}/k v_T)^{1/2}$. This constraint is less strict for temperatures $T_b/T > (\gamma_{th}/k v_T)^{2}$, where the critical density increases to $n_b/n_0 \sim (T_b/T)^{1/4}$. A high $\sigma_c$ is not strictly enough to conclude that the background can be neglected. However, it does imply a low $n_b/n_0$ even if the Harris temperature is moderately relativistic.
We have shown that, while a small background was not enough to affect the linear tearing growth rate, it limits the fast-growing nonlinear growth rate to the prediction for $a/\rho_L=a/d_{e,C}(n_b)\approx (a/\rho_{L,C}) \sqrt{n_b/n_0}$.

\section*{Acknowledgements}
This work is supported by the German Science Foundation DFG within the Collaborative Research Center SFB1491. The authors gratefully acknowledge the Gauss Centre for Supercomputing (GCS) e.V. (www.gauss-center.eu) for funding this project by providing computing time on the GCS Supercomputer SuperMUC-NG at Leibniz Supercomputing Centre (www.lrz.de) through the projects “Heat flux regulation by collisionless processes in heliospheric plasmas—ARIEL” and "Investigation of suprathermal features in the velocity distribution functions of space and astrophysical plasmas-SupraSpace".

\bibliographystyle{jpp}

\bibliography{pairtearing}

\begin{thebibliography}{47}
\expandafter\ifx\csname natexlab\endcsname\relax\def\natexlab#1{#1}\fi
\def\au#1{#1} \def\ed#1{#1} \def\yr#1{#1}\def\at#1{#1}\def\jt#1{\textit{#1}} \def\bt#1{#1}\def\bvol#1{\textbf{#1}} \def\vol#1{#1} \def\pg#1{#1} \def\publ#1{#1}\def\arxiv#1{#1}\def\org#1{#1}\def\st#1{\textit{#1}}

\bibitem[Barkov \& Komissarov(2016)]{Barkov2016}
{\sc \au{Barkov, Maxim~V.} \& \au{Komissarov, Serguei~S.}} \yr{2016}  \at{{Relativistic tearing and drift-kink instabilities in two-fluid simulations}}.  \jt{Monthly Notices of the Royal Astronomical Society}  \bvol{458}~(2),  \pg{1939--1947},  \arxiv{arXiv: https://academic.oup.com/mnras/article-pdf/458/2/1939/18240261/stw384.pdf}.

\bibitem[Bessho \& Bhattacharjee(2007)]{Bessho2007}
{\sc \au{Bessho, Naoki} \& \au{Bhattacharjee, A.}} \yr{2007}  \at{{Fast collisionless reconnection in electron-positron plasmas)}}.  \jt{Physics of Plasmas}  \bvol{14}~(5),  \pg{056503},  \arxiv{arXiv: https://pubs.aip.org/aip/pop/article-pdf/doi/10.1063/1.2714020/13887703/056503\_1\_online.pdf}.

\bibitem[Bessho \& Bhattacharjee(2012)]{Bessho2012}
{\sc \au{Bessho, Naoki} \& \au{Bhattacharjee, A.}} \yr{2012}  \at{Fast magnetic reconnection and particle acceleration in relativistic low-density electron–positron plasmas without guide field}.  \jt{The Astrophysical Journal}  \bvol{750}~(2),  \pg{129}.

\bibitem[Betar {\em et~al.\/}(2022)Betar, Del~Sarto, Ottaviani \& Ghizzo]{Betar2022}
{\sc \au{Betar, H.}, \au{Del~Sarto, D.}, \au{Ottaviani, M.} \& \au{Ghizzo, A.}} \yr{2022}  \at{Microscopic scales of linear tearing modes: a tutorial on boundary layer theory for magnetic reconnection}.  \jt{Journal of Plasma Physics}  \bvol{88}~(6),  \pg{925880601}.

\bibitem[Brittnacher {\em et~al.\/}(1995)Brittnacher, Quest \& Karimabadi]{Brittnacher1995}
{\sc \au{Brittnacher, M.}, \au{Quest, K.~B.} \& \au{Karimabadi, H.}} \yr{1995}  \at{A new approach to the linear theory of single-species tearing in two-dimensional quasi-neutral sheets}.  \jt{Journal of Geophysical Research: Space Physics}  \bvol{100}~(A3),  \pg{3551--3562},  \arxiv{arXiv: https://agupubs.onlinelibrary.wiley.com/doi/pdf/10.1029/94JA02743}.

\bibitem[Burkhart \& Chen(1989)]{Burkhart1989}
{\sc \au{Burkhart, G.~R.} \& \au{Chen, J.}} \yr{1989}  \at{{Collisionless tearing instability of a bi‐Maxwellian neutral sheet: An integrodifferential treatment with exact particle orbits}}.  \jt{Physics of Fluids B: Plasma Physics}  \bvol{1}~(8),  \pg{1578--1588},  \arxiv{arXiv: https://pubs.aip.org/aip/pfb/article-pdf/1/8/1578/12601256/1578\_1\_online.pdf}.

\bibitem[Cassak(2011)]{Cassak2011}
{\sc \au{Cassak, P.~A.}} \yr{2011}  \at{{Theory and simulations of the scaling of magnetic reconnection with symmetric shear flow}}.  \jt{Physics of Plasmas}  \bvol{18}~(7),  \pg{072106},  \arxiv{arXiv: https://pubs.aip.org/aip/pop/article-pdf/doi/10.1063/1.3602859/15865242/072106\_1\_online.pdf}.

\bibitem[Cerutti {\em et~al.\/}(2014)Cerutti, Werner, Uzdensky \& Begelman]{Cerutti2014}
{\sc \au{Cerutti, B.}, \au{Werner, G.~R.}, \au{Uzdensky, D.~A.} \& \au{Begelman, M.~C.}} \yr{2014}  \at{Three-dimensional relativistic pair plasma reconnection with radiative feedback in the crab nebula}.  \jt{Astrophys. J.}  \bvol{782}~(2),  \pg{104}.

\bibitem[Chen \& Lee(1985)]{Chen1985}
{\sc \au{Chen, J.} \& \au{Lee, Y.~C.}} \yr{1985}  \at{{Collisionless tearing instability in a non‐Maxwellian neutral sheet: An integrodifferential formulation}}.  \jt{The Physics of Fluids}  \bvol{28}~(7),  \pg{2137--2146},  \arxiv{arXiv: https://pubs.aip.org/aip/pfl/article-pdf/28/7/2137/12617484/2137\_1\_online.pdf}.

\bibitem[Comisso \& Sironi(2018)]{Comisso2018}
{\sc \au{Comisso, Luca} \& \au{Sironi, Lorenzo}} \yr{2018}  \at{Particle acceleration in relativistic plasma turbulence}.  \jt{Physical review letters}  \bvol{121}~(25),  \pg{255101}.

\bibitem[Coppi {\em et~al.\/}(1966)Coppi, Laval \& Pellat]{Coppi1966}
{\sc \au{Coppi, B.}, \au{Laval, G.} \& \au{Pellat, R.}} \yr{1966}  \at{Dynamics of the geomagnetic tail}.  \jt{Phys. Rev. Lett.}  \bvol{16},  \pg{1207--1210}.

\bibitem[Daughton(1999)]{Daughton1999}
{\sc \au{Daughton, William}} \yr{1999}  \at{{The unstable eigenmodes of a neutral sheet}}.  \jt{Physics of Plasmas}  \bvol{6}~(4),  \pg{1329--1343},  \arxiv{arXiv: https://pubs.aip.org/aip/pop/article-pdf/6/4/1329/19276577/1329\_1\_online.pdf}.

\bibitem[Daughton(2003)]{Daughton2003}
{\sc \au{Daughton, William}} \yr{2003}  \at{{Electromagnetic properties of the lower-hybrid drift instability in a thin current sheet}}.  \jt{Physics of Plasmas}  \bvol{10}~(8),  \pg{3103--3119},  \arxiv{arXiv: https://pubs.aip.org/aip/pop/article-pdf/10/8/3103/19271065/3103\_1\_online.pdf}.

\bibitem[Daughton \& Karimabadi(2005)]{Daughton2005}
{\sc \au{Daughton, William} \& \au{Karimabadi, H.}} \yr{2005}  \at{Kinetic theory of collisionless tearing at the magnetopause}.  \jt{Journal of Geophysical Research: Space Physics}  \bvol{110}~(A3),  \arxiv{arXiv: https://agupubs.onlinelibrary.wiley.com/doi/pdf/10.1029/2004JA010751}.

\bibitem[Del~Sarto {\em et~al.\/}(2016)Del~Sarto, Pucci, Tenerani \& Velli]{DelSarto2016}
{\sc \au{Del~Sarto, Daniele}, \au{Pucci, Fulvia}, \au{Tenerani, Anna} \& \au{Velli, Marco}} \yr{2016}  \at{“ideal” tearing and the transition to fast reconnection in the weakly collisional mhd and emhd regimes}.  \jt{Journal of Geophysical Research: Space Physics}  \bvol{121}~(3),  \pg{1857--1873},  \arxiv{arXiv: https://agupubs.onlinelibrary.wiley.com/doi/pdf/10.1002/2015JA021975}.

\bibitem[Drake \& Lee(1977)]{Drake1977}
{\sc \au{Drake, J.~F.} \& \au{Lee, Y.~C.}} \yr{1977}  \at{{Kinetic theory of tearing instabilities}}.  \jt{The Physics of Fluids}  \bvol{20}~(8),  \pg{1341--1353},  \arxiv{arXiv: https://pubs.aip.org/aip/pfl/article-pdf/20/8/1341/12612734/1341\_1\_online.pdf}.

\bibitem[{Eichmann} {\em et~al.\/}(2022){Eichmann}, {Oikonomou}, {Salvatore}, {Dettmar} \& {Tjus}]{Eichmann+2022}
{\sc \au{{Eichmann}, Bj{\"o}rn}, \au{{Oikonomou}, Foteini}, \au{{Salvatore}, Silvia}, \au{{Dettmar}, Ralf-J{\"u}rgen} \& \au{{Tjus}, Julia~Becker}} \at{ \yr{2022} } \jt{Astrophys. J.}  \bvol{939}~(1),  \pg{43},  \arxiv{arXiv: 2207.00102}.

\bibitem[Faganello {\em et~al.\/}(2010)Faganello, Pegoraro, Califano \& Marradi]{Faganello2010}
{\sc \au{Faganello, M.}, \au{Pegoraro, F.}, \au{Califano, F.} \& \au{Marradi, L.}} \yr{2010}  \at{{Collisionless magnetic reconnection in the presence of a sheared velocity field}}.  \jt{Physics of Plasmas}  \bvol{17}~(6),  \pg{062102},  \arxiv{arXiv: https://pubs.aip.org/aip/pop/article-pdf/doi/10.1063/1.3430640/16029807/062102\_1\_online.pdf}.

\bibitem[Fonseca {\em et~al.\/}(2002)Fonseca, Silva, Tsung, Decyk, Lu, Ren, Mori, Deng, Lee, Katsouleas \& Adam]{OSIRIS}
{\sc \au{Fonseca, R.~A.}, \au{Silva, L.~O.}, \au{Tsung, F.~S.}, \au{Decyk, V.~K.}, \au{Lu, W.}, \au{Ren, C.}, \au{Mori, W.~B.}, \au{Deng, S.}, \au{Lee, S.}, \au{Katsouleas, T.} \& \au{Adam, J.~C.}} \yr{2002} {\em {OSIRIS: A three-dimensional, fully relativistic particle in cell code for modeling plasma based accelerators}\/}, ,  \vol{vol. 2331}.  \publ{Springer Berlin / Heidelberg}.

\bibitem[Furth {\em et~al.\/}(1963)Furth, Killeen \& Rosenbluth]{Furth1963}
{\sc \au{Furth, Harold~P.}, \au{Killeen, John} \& \au{Rosenbluth, Marshall~N.}} \yr{1963}  \at{{Finite‐Resistivity Instabilities of a Sheet Pinch}}.  \jt{The Physics of Fluids}  \bvol{6}~(4),  \pg{459--484},  \arxiv{arXiv: https://pubs.aip.org/aip/pfl/article-pdf/6/4/459/12485401/459\_1\_online.pdf}.

\bibitem[Galeev {\em et~al.\/}(1978)Galeev, Coroniti \& Ashour-Abdalla]{Galeev1978}
{\sc \au{Galeev, A.~A.}, \au{Coroniti, F.~V.} \& \au{Ashour-Abdalla, M.}} \yr{1978}  \at{Explosive tearing mode reconnection in the magnetospheric tail}.  \jt{Geophysical Research Letters}  \bvol{5}~(8),  \pg{707--710},  \arxiv{arXiv: https://agupubs.onlinelibrary.wiley.com/doi/pdf/10.1029/GL005i008p00707}.

\bibitem[Harris(1962)]{Harris1962}
{\sc \au{Harris, E.~G.}} \yr{1962}  \at{On a plasma sheath separating regions of oppositely directed magnetic field}.  \jt{Il Nuovo Cimento (1955-1965)}  \bvol{23}~(1),  \pg{115--121}.

\bibitem[{Hoshino}(2020)]{Hoshino2020}
{\sc \au{{Hoshino}, M.}} \yr{2020}  \at{{Stabilization of Magnetic Reconnection in the Relativistic Current Sheet}}.  \jt{Astr. Phys. Jour.}  \bvol{900}~(1),  \pg{66},  \arxiv{arXiv: 2006.15501}.

\bibitem[Hoshino(2021)]{Hoshino2021}
{\sc \au{Hoshino, Masahiro}} \yr{2021}  \at{{Nonlinear explosive magnetic reconnection in a collisionless system}}.  \jt{Physics of Plasmas}  \bvol{28}~(6),  \pg{062106},  \arxiv{arXiv: https://pubs.aip.org/aip/pop/article-pdf/doi/10.1063/5.0050389/13313597/062106\_1\_online.pdf}.

\bibitem[{IceCube Collaboration}(2022)]{IceCube2022_ngc1068}
{\sc \au{{IceCube Collaboration}}} \at{ \yr{2022} } \jt{Science}  \bvol{378}~(6619),  \pg{538--543},  \arxiv{arXiv: https://www.science.org/doi/pdf/10.1126/science.abg3395}.

\bibitem[Innocenti {\em et~al.\/}(2015)Innocenti, Goldman, Newman, Markidis \& Lapenta]{Innocenti2015}
{\sc \au{Innocenti, M.~E.}, \au{Goldman, M.}, \au{Newman, D.}, \au{Markidis, S.} \& \au{Lapenta, G.}} \yr{2015}  \at{Evidence of magnetic field switch-off in collisionless magnetic reconnection}.  \jt{The Astrophysical Journal Letters}  \bvol{810}~(2),  \pg{L19}.

\bibitem[{Inoue} {\em et~al.\/}(2020){Inoue}, {Khangulyan} \& {Doi}]{Inoue+2020}
{\sc \au{{Inoue}, Yoshiyuki}, \au{{Khangulyan}, Dmitry} \& \au{{Doi}, Akihiro}} \at{ \yr{2020} } \jt{Astrophys. J. Lett.}  \bvol{891}~(2),  \pg{L33}.

\bibitem[{Kheirandish} {\em et~al.\/}(2021){Kheirandish}, {Murase} \& {Kimura}]{Kheirandish+2021}
{\sc \au{{Kheirandish}, Ali}, \au{{Murase}, Kohta} \& \au{{Kimura}, Shigeo~S.}} \yr{2021}  \at{{High-energy Neutrinos from Magnetized Coronae of Active Galactic Nuclei and Prospects for Identification of Seyfert Galaxies and Quasars in Neutrino Telescopes}}.  \jt{Astrophys. J.}  \bvol{922}~(1),  \pg{45},  \arxiv{arXiv: 2102.04475}.

\bibitem[Kirk \& Skj{\ae}raasen(2003)]{KirkHarris}
{\sc \au{Kirk, J.~G.} \& \au{Skj{\ae}raasen, O.}} \yr{2003}  \at{Dissipation in poynting-flux-dominated flows: The $\sigma$-problem of the crab pulsar wind}.  \jt{Astr. Phys. Jour.}  \bvol{591}~(1),  \pg{366}.

\bibitem[Komissarov {\em et~al.\/}(2006)Komissarov, Barkov \& Lyutikov]{Komissarov2006}
{\sc \au{Komissarov, S.~S.}, \au{Barkov, M.} \& \au{Lyutikov, M.}} \yr{2006}  \at{{Tearing instability in relativistic magnetically dominated plasmas}}.  \jt{Monthly Notices of the Royal Astronomical Society}  \bvol{374}~(2),  \pg{415--426},  \arxiv{arXiv: https://academic.oup.com/mnras/article-pdf/374/2/415/4864180/mnras0374-0415.pdf}.

\bibitem[Laval {\em et~al.\/}(1966)Laval, Pellat \& Vuillemin]{Laval1966}
{\sc \au{Laval, G.}, \au{Pellat, R.} \& \au{Vuillemin, M.}} \yr{1966} {\em {Electromagnetic Instabilities in a Collisionless Plasma}\/}, ,  \vol{vol.~2}.  \publ{International Atomic Energy Agency (IAEA)}.

\bibitem[Liu {\em et~al.\/}(2012)Liu, Drake \& Swisdak]{Liu2012}
{\sc \au{Liu, Yi-Hsin}, \au{Drake, J.~F.} \& \au{Swisdak, M.}} \yr{2012}  \at{{The structure of the magnetic reconnection exhaust boundary}}.  \jt{Physics of Plasmas}  \bvol{19}~(2),  \pg{022110},  \arxiv{arXiv: https://pubs.aip.org/aip/pop/article-pdf/doi/10.1063/1.3685755/13715521/022110\_1\_online.pdf}.

\bibitem[Liu {\em et~al.\/}(2015)Liu, Guo, Daughton, Li \& Hesse]{Liu2015}
{\sc \au{Liu, Yi-Hsin}, \au{Guo, Fan}, \au{Daughton, William}, \au{Li, Hui} \& \au{Hesse, Michael}} \yr{2015}  \at{Scaling of magnetic reconnection in relativistic collisionless pair plasmas}.  \jt{Phys. Rev. Lett.}  \bvol{114},  \pg{095002}.

\bibitem[Mbarek {\em et~al.\/}(2022)Mbarek, Haggerty, Sironi, Shay \& Caprioli]{Mbarek2022}
{\sc \au{Mbarek, Rostom}, \au{Haggerty, Colby}, \au{Sironi, Lorenzo}, \au{Shay, Michael} \& \au{Caprioli, Damiano}} \yr{2022}  \at{Relativistic asymmetric magnetic reconnection}.  \jt{Phys. Rev. Lett.}  \bvol{128},  \pg{145101}.

\bibitem[Pritchett {\em et~al.\/}(1991)Pritchett, Coroniti, Pellat \& Karimabadi]{Pritchett1991}
{\sc \au{Pritchett, P.~L.}, \au{Coroniti, F.~V.}, \au{Pellat, R.} \& \au{Karimabadi, H.}} \yr{1991}  \at{Collisionless reconnection in two-dimensional magnetotail equilibria}.  \jt{Journal of Geophysical Research: Space Physics}  \bvol{96}~(A7),  \pg{11523--11538},  \arxiv{arXiv: https://agupubs.onlinelibrary.wiley.com/doi/pdf/10.1029/91JA01094}.

\bibitem[Pucci {\em et~al.\/}(2018{\natexlab{{\em a\/}}})Pucci, Usami, Ji, Guo, Horiuchi, Okamura, Fox, Jara-Almonte, Yamada \& Yoo]{Pucci2018b}
{\sc \au{Pucci, F.}, \au{Usami, S.}, \au{Ji, H.}, \au{Guo, X.}, \au{Horiuchi, R.}, \au{Okamura, S.}, \au{Fox, W.}, \au{Jara-Almonte, J.}, \au{Yamada, M.} \& \au{Yoo, J.}} \yr{2018{\natexlab{{\em a\/}}}}  \at{{Energy transfer and electron energization in collisionless magnetic reconnection for different guide-field intensities}}.  \jt{Physics of Plasmas}  \bvol{25}~(12),  \pg{122111},  \arxiv{arXiv: https://pubs.aip.org/aip/pop/article-pdf/doi/10.1063/1.5050992/15778141/122111\_1\_online.pdf}.

\bibitem[Pucci {\em et~al.\/}(2018{\natexlab{{\em b\/}}})Pucci, Velli, Tenerani \& Del~Sarto]{Pucci2018}
{\sc \au{Pucci, F.}, \au{Velli, M.}, \au{Tenerani, A.} \& \au{Del~Sarto, D.}} \yr{2018{\natexlab{{\em b\/}}}}  \at{{Onset of fast “ideal” tearing in thin current sheets: Dependence on the equilibrium current profile}}.  \jt{Physics of Plasmas}  \bvol{25}~(3),  \pg{032113},  \arxiv{arXiv: https://pubs.aip.org/aip/pop/article-pdf/doi/10.1063/1.5022988/16153604/032113\_1\_online.pdf}.

\bibitem[Pétri \& Kirk(2007)]{Petri2007}
{\sc \au{Pétri, J} \& \au{Kirk, J~G}} \yr{2007}  \at{Growth rates of the weibel and tearing mode instabilities in a relativistic pair plasma}.  \jt{Plasma Physics and Controlled Fusion}  \bvol{49}~(11),  \pg{1885}.

\bibitem[{Tenerani} {\em et~al.\/}(2016){Tenerani}, {Velli}, {Pucci}, {Landi} \& {Rappazzo}]{2016JPlPh..82e5301T}
{\sc \au{{Tenerani}, A.}, \au{{Velli}, M.}, \au{{Pucci}, F.}, \au{{Landi}, S.} \& \au{{Rappazzo}, A.~F.}} \yr{2016}  \at{{`Ideally' unstable current sheets and the triggering of fast magnetic reconnection}}.  \jt{Journal of Plasma Physics}  \bvol{82}~(5),  \pg{535820501},  \arxiv{arXiv: 1608.05066}.

\bibitem[Yang(2017)]{Yang2017}
{\sc \au{Yang, Shu-Di}} \yr{2017}  \at{{Complete energy balance relation in relativistic magnetic reconnection and its application for guide-field reconnection}}.  \jt{Physics of Plasmas}  \bvol{24}~(1),  \pg{012904},  \arxiv{arXiv: https://pubs.aip.org/aip/pop/article-pdf/doi/10.1063/1.4973833/16130988/012904\_1\_online.pdf}.

\bibitem[Yang(2019{\natexlab{{\em a\/}}})]{Yang2019b}
{\sc \au{Yang, S.~D.}} \yr{2019{\natexlab{{\em a\/}}}}  \at{Relativistic plasmoid instability in pair plasmas}.  \jt{The Astrophysical Journal}  \bvol{882}~(2),  \pg{105}.

\bibitem[Yang(2019{\natexlab{{\em b\/}}})]{Yang2019}
{\sc \au{Yang, S.~D.}} \yr{2019{\natexlab{{\em b\/}}}}  \at{Relativistic tearing mode in pair plasmas and application to magnetic giant flares}.  \jt{The Astrophysical Journal}  \bvol{880}~(1),  \pg{44}.

\bibitem[Yin {\em et~al.\/}(2008)Yin, Daughton, Karimabadi, Albright, Bowers \& Margulies]{Yin2008}
{\sc \au{Yin, L.}, \au{Daughton, W.}, \au{Karimabadi, H.}, \au{Albright, B.~J.}, \au{Bowers, Kevin~J.} \& \au{Margulies, J.}} \yr{2008}  \at{Three-dimensional dynamics of collisionless magnetic reconnection in large-scale pair plasmas}.  \jt{Phys. Rev. Lett.}  \bvol{101},  \pg{125001}.

\bibitem[Zelenyi \& Krasnosel'skikh(1979)]{Zelenyi1979}
{\sc \au{Zelenyi, L.~M.} \& \au{Krasnosel'skikh, V.~V.}} \yr{1979}  \at{Relativistic modes of tearing instability in a background plasma.}  \jt{Astron. Zh.}  \bvol{56},  \pg{819--832}.

\bibitem[Zenitani(2017)]{Zenitani2017}
{\sc \au{Zenitani, Seiji}} \yr{2017}  \at{Dissipation in relativistic pair-plasma reconnection: revisited}.  \jt{Plasma Physics and Controlled Fusion}  \bvol{60}~(1),  \pg{014028}.

\bibitem[Zenitani \& Hoshino(2007)]{Zenitani2007}
{\sc \au{Zenitani, S.} \& \au{Hoshino, M.}} \yr{2007}  \at{Particle acceleration and magnetic dissipation in relativistic current sheet of pair plasmas}.  \jt{The Astrophysical Journal}  \bvol{670}~(1),  \pg{702}.

\bibitem[Zenitani \& Hoshino(2008)]{Zenitani2008}
{\sc \au{Zenitani, S.} \& \au{Hoshino, M.}} \yr{2008}  \at{The role of the guide field in relativistic pair plasma reconnection}.  \jt{The Astrophysical Journal}  \bvol{677}~(1),  \pg{530}.

\end{thebibliography}

\end{document}